\documentclass[journal]{IEEEtran}
\usepackage{booktabs}
\usepackage{amssymb}
\usepackage{amsmath}
\usepackage{amsfonts}
\usepackage{multirow}
\usepackage{graphicx}
\usepackage{textcomp}
\usepackage{amsthm}

\usepackage{setspace}
\usepackage{subfigure}
\usepackage{algorithmic}
\usepackage[ruled,vlined,boxed,linesnumbered]{algorithm2e}
\usepackage[dvipsnames]{xcolor}
\usepackage{bm}
\usepackage{hyperref}
\usepackage{url}
\usepackage{xcolor} 

\allowdisplaybreaks[4]

\ifCLASSINFOpdf
\else
\fi

\begin{document}

\title{Hybrid Learning for Cold-Start-Aware Microservice Scheduling in Dynamic Edge Environments}

\author{Jingxi Lu,
        Wenhao Li,
        Jianxiong Guo,~\IEEEmembership{Member,~IEEE},
        Xingjian Ding,
        Zhiqing Tang,~\IEEEmembership{Member,~IEEE},
        Tian Wang,~\IEEEmembership{Senior Member,~IEEE}, and
        Weijia Jia,~\IEEEmembership{Fellow,~IEEE}
	
    \thanks{This work is an extended version of the paper \cite{lu2024Container}, which has been published at the 20th International Conference on Mobility, Sensing and Networking (MSN 2024), Harbin, China.
    
    Jingxi Lu is with the Ming Hsieh Department of Electrical and Computer Engineering, University of Southern California, Los Angeles, CA 90089, USA. (E-mail: jingxil@usc.edu)
    
    Wenhao Li is with the School of Computer Science, The University of Sydney, Sydney,  NSW 2037, Australia. (E-mail: weli0278@uni.sydney.edu.au)

    Jianxiong Guo and Weijia Jia are with the Advanced Institute of Natural Sciences, Beijing Normal University, Zhuhai 519087, China, and also with the Guangdong Key Lab of AI and Multi-Modal Data Processing, Beijing Normal-Hong Kong Baptist University, Zhuhai 519087, China. (E-mail: \{jianxiongguo, jiawj\}@bnu.edu.cn)

    Xingjian Ding is with the Faculty of Information Technology, Beijing University of Technology, Beijing 100124, China. (E-mail: dxj@bjut.edu.cn)

    Zhiqing Tang and Tian Wang are with the Advanced Institute of Natural Sciences, Beijing Normal University, Zhuhai 519087, China. (E-mail: \{zhiqingtang, tianwang\}@bnu.edu.cn)
	
    \textit{(Corresponding author: Jianxiong Guo.)}
	}
    \thanks{Manuscript received xxxx; revised xxxx.}}

\markboth{}%
{Shell \MakeLowercase{\textit{et al.}}: Bare Demo of IEEEtran.cls for IEEE Journals}

\maketitle

\begin{abstract}    
    With the rapid growth of IoT devices and their diverse workloads, container-based microservices deployed at edge nodes have become a lightweight and scalable solution. However, existing microservice scheduling algorithms often assume static resource availability, which is unrealistic when multiple containers are assigned to an edge node. Besides, containers suffer from cold-start inefficiencies during early-stage training in currently popular reinforcement learning (RL) algorithms. In this paper, we propose a hybrid learning framework that combines offline imitation learning (IL) with online Soft Actor-Critic (SAC) optimization to enable a cold-start-aware microservice scheduling with dynamic allocation for computing resources. We first formulate a delay-and-energy-aware scheduling problem and construct a rule-based expert to generate demonstration data for behavior cloning. Then, a GRU-enhanced policy network is designed in the policy network to extract the correlation among multiple decisions by separately encoding slow-evolving node states and fast-changing microservice features, and an action selection mechanism is given to speed up the convergence. Extensive experiments show that our method significantly accelerates convergence and achieves superior final performance. Compared with baselines, our algorithm improves the total objective by $50\%$ and convergence speed by $70\%$, and demonstrates the highest stability and robustness across various edge configurations.
\end{abstract}


\IEEEpeerreviewmaketitle

\section{Introduction}
With the rapid advancement of Artificial Intelligence (AI), Internet of Things (IoT), and the Fifth Generation Mobile Network (5G), the number of connected edge devices and the volume of data they generate have grown exponentially. Traditional cloud computing offloads such data to centralized data centers for processing. However, this approach introduces high transmission delays and bandwidth pressure, making it unsuitable for latency-sensitive and computation-intensive applications \cite{khan2019edge}. To address these issues, \textit{Mobile Edge Computing (MEC)} has emerged as a promising paradigm that brings computational capabilities closer to data sources.

Meanwhile, the \textit{microservice architecture} has gained popularity in cloud-native application design for its modularity, scalability, and ease of deployment \cite{wolff2016microservices,nadareishvili2016microservice}. Containers, due to their lightweight virtualization capabilities, are widely used to encapsulate and deploy microservices efficiently \cite{zhou2018scheduling}. Several practical container scheduling solutions have been proposed to deploy microservices in MEC scenarios. However, most widely-used orchestration platforms, such as Docker \cite{docker} and Kubernetes \cite{kubernetes}, adopt simplistic strategies that mainly consider static resource availability and do not support intelligent, context-aware decision-making \cite{fazio2016open}.

More recent research has begun to consider advanced objectives such as delay minimization, load balancing, and image reuse \cite{ma2018efficient,wang2019delay}. However, many of these models still rely on unrealistic assumptions. For instance, they often assume that the computing capacity on edge nodes is constant and evenly distributed across tasks \cite{Cui_Tang_Lou_Jia_2023}, or that tasks are processed sequentially via queuing mechanisms, which may introduce significant latency. These assumptions limit the practicality of existing models, especially when edge nodes must handle concurrent microservices under dynamic and constrained resources.

The primary scheduling challenge lies in allocating multiple microservices to multiple edge nodes, considering the distribution of container images and the allocation of computing power when multiple microservices are assigned concurrently. Here, we adopt the Round-Robin (RR) strategy \cite{rasmussen2008round} to equally distribute the remaining computing power to all newly arriving tasks. Once it is determined, it will not change until the task is completed. This is in line with the container management mechanism in Kubernetes. Once a container is created and executed, corresponding computing resources should be allocated. However, this creates new challenges: since the number of microservices ultimately assigned to each node is unknown at scheduling time, naive decisions may lead to node overload and degraded performance.

\begin{figure}[!t]
    \centering
    \includegraphics[width=\linewidth]{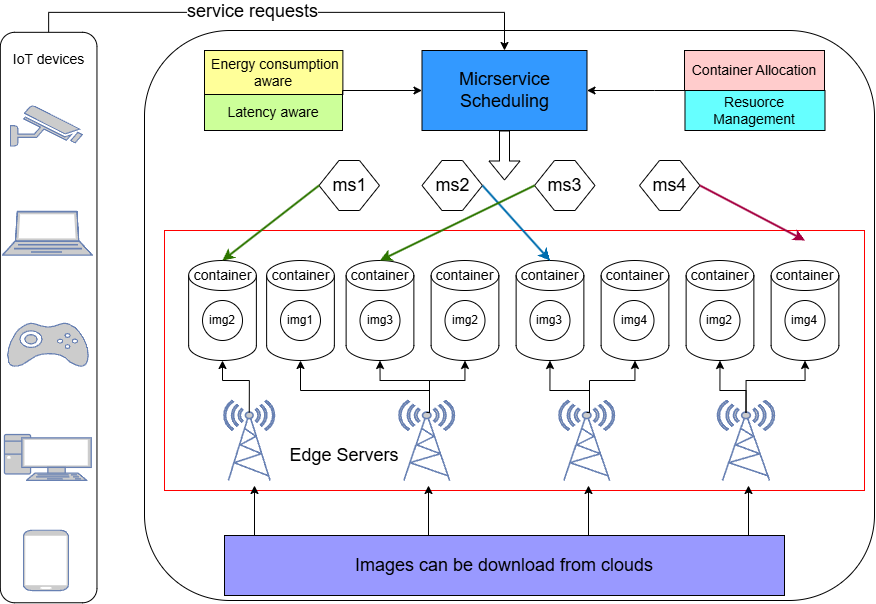}
    \caption{The microservice management architecture: Demonstrates our microservice deployment strategy using containers for optimized service delay and energy usage across edge nodes, maintaining acceptable performance under variable loads.}
    \label{layout}
\end{figure}

To address the complexity of this decision space, recent works have modeled online microservice scheduling as a Markov Decision Process (MDP) and applied Reinforcement Learning (RL) algorithms to learn adaptive scheduling policies \cite{mao2016resource}. Among these, Soft Actor-Critic (SAC) \cite{haarnoja2018soft} has shown strong performance due to its entropy-regularized policy updates and fast convergence. Thus, it is used to implement our online scheduling system due to its good performance and fast convergence. However, in our problem, there are many decisions to be made, including the assignment for all microservices arriving in a time slot, which leads to a huge space for action, and RL cannot effectively deal with a large-scale action space. It can be considered that all actions that need to be done at the same time are decided in sequence, but only the same state can be observed, and all decisions of microservices interact with each other. Therefore, in our earlier work \cite{lu2024Container}, we added a GRU unit to the policy network to extract the correlation between multiple actions at this time. The hidden state of this GRU is cleared every time slot because it only focuses on the things in this time slot.

However, online RL methods typically suffer from low sample efficiency and unstable convergence during early training, commonly known as the cold-start problem. This issue is particularly acute in our setting, where multiple microservices must be scheduled concurrently under stringent delay constraints and limited edge resources. A poorly initialized policy in the early stage may assign an excessive number of tasks to a single edge node, overlook image locality, or ignore task deadlines, leading to severe resource contention, service delay violations, and energy inefficiencies. Unlike environments with stationary dynamics, microservice scheduling in edge computing involves dynamic resource profiles and strong inter-task dependencies. A suboptimal early decision can propagate across tasks and nodes, triggering cascading failures in system performance \cite{li2021joint}. These challenges make cold-start particularly detrimental in edge-based container scheduling, where every time slot decision impacts multiple co-located services.

In this paper, we propose a two-phase learning framework that combines offline Imitation Learning (IL) with online RL to tackle the problem of container-based microservice scheduling under dynamic edge resource conditions. In the offline phase, we construct a rule-based expert policy that jointly considers container image locality, resource constraints, and delay-energy trade-offs to generate expert demonstrations. These demonstrations are used to pre-train the policy network via behavior cloning, providing a strong initialization. In the online phase, the policy is fine-tuned using a GRU-enhanced SAC algorithm, which enables sequential decision-making across temporally correlated microservices with dynamic CPU allocation. We implement the proposed approach in a customized container scheduling simulator, with source code available at \url{https://github.com/Blacktower27/CSDCRMDE}. Experimental results show that our method significantly reduces service latency and energy consumption while achieving fast and stable convergence, consistently outperforming existing baselines. The main contributions of this journal extension are as follows:

\begin{itemize}
    \item  To the best of our knowledge, we are the first to mathematically model the online container-based microservice scheduling problem with dynamic computing power to minimize total latency and energy consumption, which improves the reliability and efficiency of microservice applications in resource-constrained edge nodes.
    \item We extend our earlier work \cite{lu2024Container} by introducing a two-phase learning framework that combines offline imitation learning with online reinforcement learning. The offline phase uses a rule-based expert policy to generate demonstration data, effectively addressing the cold-start issue and improving training stability.
    \item We propose an enhanced policy network that decouples state components by temporal characteristics: a GRU encoder captures accumulated scheduling context from slowly evolving node states, while a linear layer models fast-changing microservice states. This architecture enables scalable, context-aware multi-task scheduling.
    \item Extensive experiments demonstrate that our model achieves the highest final reward and the fastest, most stable convergence among all baselines. Specifically, it improves the total objective by over 50\% and reduces the convergence time by nearly 70\% compared to standard SAC and PPO variants. As shown in Fig.~\ref{rewardcomp}, it consistently outperforms both fully connected and purely online learning approaches, confirming the effectiveness of our two-phase GRU-based enhanced design.
\end{itemize}

\textbf{Organization.} We first summarize the works related to this paper in Section \ref{sec2}. Section \ref{sec3} presents the system model and formulates the online microservice scheduling problem under dynamic computing resources. Section \ref{sec4} introduces the design of our two-phase algorithm along with the details of the enhanced network architecture. Section \ref{sec5} provides experimental comparisons and performance evaluation. Finally, Section \ref{sec6} concludes the paper.

\section{Related Work}\label{sec2}
We categorize related studies into two key perspectives: (1) system-level design for container-based microservice scheduling in dynamic edge environments, and (2) learning-based resource scheduling, including recent advances in IL and Deep Reinforcement Learning (DRL).
\subsection{Container-based Microservice Scheduling}
With the rise of cloud-native modularization, containers have become the standard for deploying microservices due to their lightweight virtualization, fast startup, and isolation capabilities \cite{zhou2018scheduling, chung2018stratus}. Compared to traditional virtual machines (VMs), containerized deployment simplifies runtime resource management by sharing the host operating system kernel and reducing memory footprints. Early solutions focused on basic heuristics and static deployment rules. Singh \textit{et al.} \cite{singh2017container} automated microservice deployment using Docker, while Wan \textit{et al.} \cite{wan2018application} and Abdullah \textit{et al.} \cite{ABDULLAH2019243} explored application-layer decomposition and orchestration strategies. However, these methods struggle in highly dynamic and resource-constrained edge environments.

To address this, system-level works began incorporating robustness, reuse, and spatio-temporal modeling in this area. Chen \textit{et al.} \cite{chen2018spatio} treated edge service placement as a contextual bandit problem, capturing spatial and temporal variations. Li \textit{et al.} \cite{li2020read} developed a robustness-aware scheduling framework, and Deng \textit{et al.} \cite{deng2020optimal} modeled constrained resource-aware deployment across distributed edges. Gu \textit{et al.} \cite{gu2021layer} proposed a layer-sharing strategy for co-located containers, while Fu \textit{et al.} \cite{fu2021adaptive} designed Nautilus, a runtime system for latency-sensitive workloads.

\subsection{Learning-based Scheduling: DRL and IL}
RL has emerged as a promising approach for online container scheduling, as it enables adaptive decision-making in uncertain and time-varying environments. Wang \textit{et al.} \cite{wang2019delay} and Chen \textit{et al.} \cite{chen2020iot} applied DQN and DDPG for delay-aware container orchestration. Lv \textit{et al.} \cite{lv2022microservice, lv2023graph} proposed a graph-based deep Q-learning strategy based on reward sharing for dependency-aware scheduling. Tan \textit{et al.} \cite{tan2020nsga} integrated NSGA-II with actor-critic RL to address trade-offs between latency and energy. Shen \textit{et al.} \cite{shen2023collaborative} further proposed a collaborative learning-based scheduler tailored for Kubernetes in edge-cloud networks, capturing workload dynamics in a distributed manner. Beyond container-centric orchestration, Luo \textit{et al.} \cite{luo2020collaborative} and Liu \textit{et al.} \cite{liu2019drl} employed DRL to optimize task offloading and data scheduling in vehicular edge environments, demonstrating its adaptability to complex spatiotemporal resource dynamics.

Despite the success of DRL in adaptive scheduling, existing methods still face challenges in terms of convergence efficiency and sensitivity to dynamic workloads. To improve early-stage performance and reduce inefficient exploration, recent studies have explored the integration of IL into the scheduling pipeline. Given the complexity and dynamic nature of microservice environments, rule-based heuristics or random initialization often provide insufficient prior knowledge, making IL particularly appealing due to its ability to incorporate expert demonstrations. Although IL remains underutilized in container orchestration, systems such as DL2 \cite{peng2021dl2} have demonstrated the benefits of leveraging historical expert data for efficient job scheduling. Hussein \textit{et al.} \cite{hussein2017survey} further offered a comprehensive taxonomy of IL approaches, reinforcing their relevance in real-time sequential decision-making tasks like scheduling. Integrating IL with RL provides a practical framework for efficient policy initialization, mitigating instability, and accelerating convergence during early training phases.

In the context of container-based orchestration, Zhao \textit{et al.} \cite{zhao2023deep} proposed a Dueling DQN-based method for online microservice deployment in mobile edge computing, aiming to improve deployment success rate and responsiveness in dynamic environments. Sang \textit{et al.} \cite{sang2024kubernetes} designed a Kubernetes-compatible scheduling algorithm trained via IL, which achieved high deployment efficiency under heterogeneous resource constraints and limited training data. Additionally, Toka \textit{et al.} \cite{toka2021machine} introduced a machine learning-based autoscaling engine for cloud-hosted Kubernetes services, which dynamically selects forecasting models to better match provisioning decisions to workload demand. 

\subsection{Discussion}
While prior works have demonstrated the potential of applying DRL and IL to container-based microservice scheduling, they still face fundamental limitations in dynamic environments. Existing DRL-based methods typically assume static resource conditions and treat scheduling decisions independently, thereby neglecting dynamic variations in computing resources and interdependencies between intra-slot decisions. Although IL-enhanced frameworks improve early-stage training stability, they often focus on deployment success or resource matching, without systematically addressing the joint optimization of latency, energy consumption, and resource feasibility under dynamic constraints. To overcome these challenges, we propose a hybrid imitation-reinforcement learning framework that explicitly models dynamic CPU allocation, captures sequential decision correlations through a GRU-enhanced policy network, and leverages expert demonstrations, balancing multiple scheduling objectives. This design aims to enhance adaptability, sample efficiency, and convergence stability in dynamic environments.

\section{System Model \& Problem Formulation}\label{sec3}
In this section, the system is first modeled, and then the service latency and energy consumption are defined. Next, the concept and function of the deadline of the microservice requests are introduced. Finally, we discuss the constraints, and the optimization problem is formulated.

\subsection{System model}
We first model the offloading scenario for computational microservices. As shown in Fig. \ref{layout}, in the Internet of Things (IoT) environment, computation-intensive microservice requests are generated by IoT devices and offloaded to a group of edge nodes $\mathcal{N}$ for processing. The set of nodes is defined as $\mathcal{N}=\{n_1,n_2,\cdots,n_{N}\}$, where $N$ denotes the number of nodes in this set. Each edge node $n$ possesses limited resources, primarily including total CPU frequency $F_n$, memory resource $M_n$, and storage space $D_n$, and can concurrently run a finite number of containers. The system time is divided into consecutive time slots of equal length, denoted by $\mathcal{T}=\{t_1,t_2,\cdots\}$ with the same length $\Delta$. At the beginning of each time slot $t$, a set of microservice requests $\mathcal{K}_t$ are generated by different IoT devices, and they can be denoted by $\mathcal{K}_t=\{k_1,k_2, \cdots,k_{K_t}\}$.    

Each microservice request $k$ in $\mathcal{K}_t$ should be executed within a container and requires a specific image from $\mathcal{I} =\{i_1,i_2, \cdots, i_{I}\}$ to initialize this container, which can be denoted by $k=\{d_k,c_k,m_k,i_k,l_k\}$, where $d_k$ is the size of the microservice, $c_k$ is the total number of CPU cycles required to accomplish the computation microservice, $m_k$ is the required memory of the microservice, $i_k\in\mathcal{I}$ is the required image of the microservice, and $l_k$ is the maximum delay tolerance of the request. IoT devices can communicate with the edge node via a wireless connection to determine which node the microservice should be offloaded to.

\subsection{Microservice Latency}
In the process of a demand for microservices, there will inevitably be some delays, which directly affect the performance of the system and user experience. Major delays include communication latency, image download latency, and computation latency, which are critical in IoT and edge computing environments.

\textbf{Communication latency.}
Regarding communication latency, we consider a communication model in which IoT devices share the bandwidth of edge nodes. The uplink wireless transmission rate $\xi_{n,k}(t)$ of the device of microservice request $k$ to edge node $n$ is calculated as:
\begin{equation} \xi_{n,k}(t) = \frac{B_n}{U_n(t)} \log\left(1 + \frac{p_k h_{n,k}}{\sigma^2} \right), \end{equation}
where $B_n$ is the bandwidth of edge node $n$, and $U_n(t)$ denotes the number of microservice requests transmitted to node $n$ at time slot $t$. In addition, $p_k$ is the transmission power, $h_{n,k}$ is the channel gain between the IoT device and the edge node, and $\sigma^2$ represents the power of Gaussian white noise. The communication latency of microservice request $k$ transmitted to node $n$ is defined as:
\begin{equation} 
T_{n,k}^{\text{comm}}(t) = {d_k}/{\xi_{n,k}(t)}, 
\end{equation}
where $d_k$ is the data size of microservice $k$. We assume that the communication delay $T_{n,k}^{\text{comm}}(t)$ does not exceed the duration of a time slot:
\begin{equation} 
T_{n,k}^{\text{comm}}(t) \leq \delta, \quad \forall n \in \mathcal{N},\ k \in \mathcal{K}_t. 
\end{equation}
Furthermore, we overlook the return communication latency of results, as it is relatively small compared to the microservice processing time.

\textbf{Image download latency.}
Image download delay refers to the latency incurred in obtaining container images related to microservice processing. It is calculated as:
\begin{equation}
T_{n,k}^{\text{down}}(t) = x_{n,i_k}(t) \times \left( \frac{s_{i_k}}{B_n} + T_n^{\text{queue}}(t) \right), 
\end{equation}
where $i_k$ denotes the container image required by microservice $k$, and $s_{i_k}$ is the size of that image. The binary variable $x_{n,i}(t) \in\{0,1\}$ indicates whether image $i$ is present on node $n$ at the beginning of time slot $t$, where $x_{n,i}(t)=0$ means the image is already present locally and $x_{n,i}(t)=1$ indicates it needs to be downloaded. $T_n^{\text{queue}}(t)$ denotes the queuing delay for image downloads on node $n$ at time $t$. Therefore, if the image is already stored on the node, the image download delay is set to zero.

\textbf{Computation latency.}
Each microservice is executed in an independent container, and all microservices are processed in parallel. The computation latency is given by:
\begin{equation} 
T_{n,k}^{\text{comp}}(t) = {c_k \cdot U_n(t)}/{F_n(t)}, 
\end{equation}
where $c_k$ denotes the required CPU cycles of microservice $k$, $U_n(t)$ is the number of microservices assigned to node $n$ at time $t$, and $F_n(t)$ represents the remaining CPU frequency of node $n$ at the beginning of time slot $t$. It is assumed that the remaining CPU frequency is evenly distributed among all scheduled microservices.

\subsection{Energy consumption.}
Each microservice will be offloaded on a node $n$. After processing the microservice, the node returns the calculation result. Note that we ignore the transmission energy consumption of returning the calculated results from the node because, in most cases, the data volume is small and the downlink rate is higher. The overall power consumption related to microservice $k$ can be categorized into two primary components. The first component pertains to the power consumed during the transfer of the microservice to the node. The second component accounts for the power consumed during the actual computation on the node.
Based on the time it takes to upload the microservice, the power consumption of the microservice $k$ upload can be defined as: 
\begin{equation}
    E_{n,k}^{comm}(t)={p_n^{comm}\cdot T_{n,k}^{comm}(t)}/{U_n(t)}
\end{equation}
where $p_n^{comm}$ is the total power of transmission in node $n$. When the node is processing the uploaded computing microservice, the corresponding energy consumption of the node can be expressed as: 
\begin{equation}
    E_{n,k}^{comp}(t)=\frac{p_n^{comp}\cdot T_{n,k}^{comp}(t)\cdot F_n(t)}{U_n(t)\cdot F_n},
\end{equation}
where $p_n^{comp}$ is the total power of computation in node $n$. It is not difficult to find that we adopt the strategy of equal distribution in transmission and computing power.

    
\subsection{Problem Formulation and Analysis}
For convenience, we define $y_{n,k}(t)\in\{0,1\}$ as an indicator variable. When $y_{n,k}(t)=1$, the microservice $k$ is executed on edge node $n$ at the time slot $t$; Otherwise, we have $y_{n,k}(t)=0$. For each microservice $k$, it can be assigned to at most one edge node, then we have $\sum_{n\in\mathcal{N}}y_{n,k}(t)=1$. For each node $n$, it has $U_n(t)=\sum_{k\in\mathcal{K}_t}y_{n,k}(t)$. Thus, the total latency $T_k(t)$ and total energy $E_k(t)$ to complete microservice $k$ at the time slot $t$ can be formulated as
\begin{align}
    T_k(t) &= \sum\nolimits_{n\in\mathcal{N}}y_{n,k}(t)\left[T_{n,k}^{comm}(t)+T_{n,k}^{down}(t)+T_{n,k}^{comp}(t)\right]\nonumber\\
    E_k(t) &= \sum\nolimits_{n\in\mathcal{N}}y_{n,k}(t)\left[E_{n,k}^{comm}(t)+E_{n,k}^{comp}(t)\right]. \label{eq:E_k}
\end{align}

At the beginning of time slot $t$, we denote the remaining CPU frequency of node $n$ as $F_n(t)$, the remaining memory of node $n$ as $M_n(t)$, and the remaining storage space of node $n$ as $D_n(t)$, respectively. Following this, the constraints in our model can be defined.
\begin{itemize}
    \item Delay constraint: the total delay to finish the microservice $k$ cannot exceed its maximum tolerance. That is
    \begin{equation}\label{delay}
        T_k(t)\leq l_k, \forall k\in\mathcal{K}_t.
    \end{equation}
    \item Memory constraint: The total memory on each node is limited. That is
    \begin{equation}\label{memory}
        \sum\nolimits_{k\in\mathcal{K}_t}y_{n,k}(t)\cdot m_k\leq M_n(t), \forall n\in\mathcal{N}.
    \end{equation}
    \item Storage constraint: The total storage on each node is limited. That is
    \begin{equation}\label{storage}
        \sum\nolimits_{k\in\mathcal{K}_t}y_{n,k}(t)\cdot x_{n,i_k}(t)\cdot s_{i_k}\leq D_n(t), \forall n\in\mathcal{N}.
    \end{equation}
\end{itemize}
Generally, once a microservice is finished, the memory and CPU resources it occupies will be released. However, once an image is downloaded to a node, it will always exist on this node and is allowed to be shared by multiple microservices. Thus, it implies $x_{n,i}(t')\leq x_{n,i}(t)$ if $t'\geq t$.

\begin{figure}[!t]
    \centering
    \includegraphics[width=\linewidth]{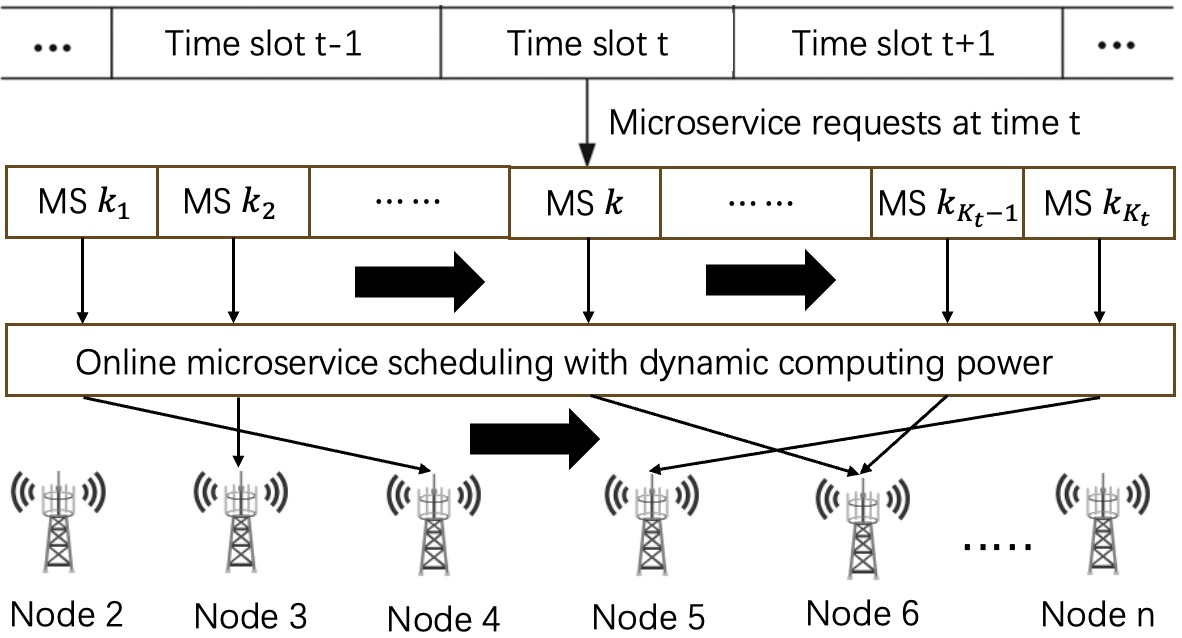}
    \caption{Overview of the system model, in each time slot $t$, there will be $K_t$ microservices that need to be offloaded to acceptable edge nodes at the same time. However, in the actual scheduling, it will be processed in a sequential manner, which will be described in Sec. \ref{sec4-2}.}
    \vspace{-10pt} 
    \label{timeslot}
\end{figure}

We aim to minimize the weighted overall microservice processing latency and energy consumption from a long-term perspective. The decision process of the microservice scheduling with dynamic computing power is shown in Fig. \ref{timeslot}. At each time slot $t$, there are totally $K_t$ microservices that should be determined at the same time. We use $\alpha$ to represent the weight assigned to latency at the time slot $t$. The target is to find the best strategy that can minimize the overall cost while obeying the constraints. Therefore, the objective of our online microservice scheduling problem can be defined as
\begin{equation}\label{prob}
    \begin{aligned}
    &\min \sum\nolimits_{t \in \mathcal{T}}\sum\nolimits_{k \in \mathcal{K}_t}\left(\alpha\cdot T_k(t) + E_k(t)\right)\\
    &\text{s.t. Constraint (\ref{delay}), (\ref{memory}), and (\ref{storage});}\\
    &\quad\enspace x_{n,i}(t)\in\{0,1\}, \forall n\in\mathcal{N}, \forall i\in\mathcal{I};\\
    &\quad\enspace x_{n,i}(t+1)\leq x_{n,i}(t);\\
    &\quad\enspace y_{n,k}(t)\in\{0,1\}, \forall n\in\mathcal{N}, \forall k\in\mathcal{K}_t, \forall t\in\mathcal{T};\\
    &\quad\enspace \sum\nolimits_{n\in\mathcal{N}}y_{n,k}(t)=1, \forall k\in\mathcal{K}_t.
    \end{aligned}
\end{equation}
Obviously, the problem defined in (\ref{prob}) is NP-hard because it is a special case of online integer programming. RL-based methods can get a good solution by continuously interacting with the environment.

\section{Algorithm Design}\label{sec4}
In this section, we will approximate the optimization problem as an MDP to minimize the energy consumption and delay during the processing of microservices. Firstly, we define the states, actions, and reward functions within the MDP framework. Secondly, we will discuss the composition of the policy network. Finally, we propose a novel intelligent computation offloading algorithm based on SAC to address the optimization problem.
\subsection{Algorithm Settings}
\textbf{State.} The system state $s^{k}_{t}$ encompasses both the node resource state and microservice state. In each time slot $t$, microservices in $\mathcal{K}_t$ are processed sequentially. The node resource state, denoted as $s^{\text{node},k}_{t}$, represents the remaining resources on each node when microservice $k\in\mathcal{K}_t$ arrives. We denote the remaining memory and storage of the node $n\in\mathcal{N}$ at time slot $t$ when microservice $k$ arrives as $M^{k}_{n}(t)$ and $D^{k}_{n}(t)$, respectively. The resource state includes the remaining CPU frequency, remaining memory capacity, storage capacity, communication power, computation power, and bandwidth of the node at time slot $t$. This state can be defined as:
\begin{align}
    s_{t}^{node,k} = \{& F_{1}(t), \ldots, F_{|N|}(t), M^{k}_{1}(t), \ldots, M^{k}_{|N|}(t),\nonumber \\
    & D^{k}_1(t), \ldots, D^{k}_{|N|}(t), P^{comm}_1, \ldots, P^{comm}_{|N|},\nonumber \\
    & P^{comp}_1, \ldots, P^{comp}_{|N|}, B_1, \ldots, B_{|N|}\}.
\end{align}
The microservice state $s_t^{ms,k}$ of microservice $k\in\mathcal{K}_t$ at time slot $t$ encompasses information about the required image on each node, $y_{n,k}(t)$ for each node $n\in\mathcal{N}$ and details about the microservice, including the required image, the required CPU cycles, the microservice size, and the maximum delay tolerance. Thus, the microservice state can be expressed as:
\begin{equation}
    s_{t}^{ms, k} = \{d_k, c_k, m_k, i_k, l_k\}.
\end{equation}
Thus, the system state $s^k_t$ is $s_{t}^{k} = s_{t}^{node,k} \cup s_{t}^{ms, k}$.

\textbf{Action space.}
The action at time $t$ for microservice $k$ is to assign it to an edge node $n$. Therefore, the action space is defined as: \begin{equation} a_{t}^{k} \in \mathcal{A} = \{1, 2, \ldots, |N|\}. \end{equation}

\textbf{Reward.}
To encourage the agent to minimize long-term energy consumption while satisfying latency constraints, we define the reward based on energy cost and delay. If the latency constraint 
\begin{equation} 
    T_{k}(t) > l_k 
\end{equation} 
is violated, the reward is set to a negative penalty. Otherwise, the reward is calculated as:
\begin{equation} 
r_{t}^{k} = \alpha \cdot (l_k - T_{k}(t)) - E_{k}(t), 
\end{equation}
where $T_k(t)$ and $E_k(t)$ denote the actual delay and energy consumption of microservice $k$ at time $t$ (computed by Eqn.~(\ref{eq:E_k})), and $\alpha$ is a weighting factor balancing latency optimization and energy saving.

\subsection{Policy Network}\label{sec4-2}
In each time slot $t$, the decision $a^k_t$ is made according to its observation $s^k_t$, however, this $s^k_t$ does not contain any information about other microservices required to be dealt with in this time slot. In other words, there are a total of $\mathcal{K}_t$ microservices that need to be addressed concurrently. For example, the remaining resources $F_n(t)$, $M_n(t)$, and $D_n(t)$ of node $n$ are sufficient at the beginning of time slot $t$. However, if too many microservices are eventually assigned to this node, it will lead to low efficiency. Thus, we need to design carefully to overcome this drawback and achieve an efficient dynamic scheduling. Since the resources of edge nodes are dynamically allocated, the CPU frequency for each microservice cannot be determined until all scheduling decisions in $\mathcal{K}_t$ are completed. During this sequential decision process, only the node memory $M^{k}_n(t)$ and storage $D^{k}_n(t)$ are updated immediately after each individual decision in the same time slot. This means that the microservice scheduling decisions made in a time slot $t$ are made sequentially, forming a time series as shown in Fig. \ref{timeslot}. This time series information destroys the Markov nature of the environment, making it impossible for the agent to generate the best action based solely on the currently observed state $s_t^k$.

For feature extraction of sequence information in microservice scheduling, we consider adding a unit of RNN-based structure in the policy network, which can learn the time dependence between states of previous microservices. By utilizing GRU, we can comprehensively integrate information from previous microservice scheduling decisions in this time slot into the current state. Thus, we can avoid making bad decisions because of the conflict with previous decisions.

\begin{figure}[!t]
    \centering
    \includegraphics[width=\linewidth]{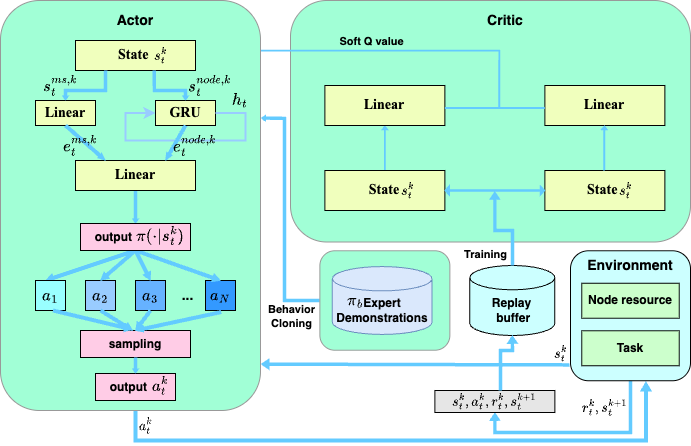}
    \caption{Overview of the SAC-based framework.}
    \label{fig1}
\end{figure}

As illustrated in Fig.~\ref{fig1}, the policy network adopts an enhanced architecture with two branches, tailored to the temporal dynamics of different state components. The node state \(s_t^{\text{node},k}\) changes slowly within a time slot. Its CPU frequency remains fixed, while other resources such as memory and storage are gradually consumed as microservices are sequentially scheduled. This low-frequency temporal behavior makes it suitable for modeling with a GRU, which captures the cumulative context across scheduling decisions and helps avoid resource overload on a single node.

To extract the sequential scheduling context, we utilize a GRU module that takes the node state $s_t^{\text{node},k}$ as input and produces two outputs: the hidden state $h_t$ and the embedding representation $e_t^{\text{node},k}$. The hidden state $h_t$ is used to maintain the temporal dependencies across microservices within the same time slot, while the embedding $e_t^{\text{node},k}$ serves as a compact representation of the accumulated scheduling context and is passed to the policy head for action prediction. 

Specifically, a GRU includes two control gates: the update gate ($z_t^k$) and the reset gate ($g_t^k$). The update gate determines how much of the previous hidden state $h_{t}$ should be preserved, while the reset gate controls how much of the previous state is considered when computing the current candidate state $\hat{h}_t$. The updated rules of the GRU are as follows:
\begin{align}
    & g_t^k = \sigma(W_{xr} \cdot s_t^{\text{node},k} + W_{hr} \cdot h_{t} + b_r), \\
    & z_t^k = \sigma(W_{xz} \cdot s_t^{\text{node},k} + W_{hz} \cdot h_{t} + b_z), \\
    & \hat{h}_t = \tanh(W_{xh} \cdot s_t^{\text{node},k} + W_{hh} \cdot (re_t^k \otimes h_{t}) + b_h), \\
    & h_t = (1 - z_t^k) \otimes h_{t} + z_t^k \otimes \hat{h}_t,\label{eq21}
\end{align}
where $W_{xr}$, $W_{hr}$, $W_{xz}$, and $W_{hz}$ are weight matrices, $b_r$ and $b_z$ are bias vectors, and $g_t^k$ and $z_t^k$ are vectors for resetting and updating the activation values of gates.

It is worth noticing that the GRU module only needs to focus on the time series feature in a single time slot. Therefore, at the beginning of each time slot, the hidden state of the GRU module $h_t$ would be reset to the same initial value.
In contrast, the microservice state \(s_t^{\text{ms},k}\) changes rapidly from one task to another due to differing resource demands, deadlines, and image requirements. To preserve its high-frequency, task-specific features, it is passed through a linear embedding layer, yielding a task representation \(e_t^{\text{ms},k}\) without temporal recurrence. The resulting hidden representations $e_t^{\text{node},k}$ and $e_t^{\text{ms},k}$ are concatenated and passed through a fully connected layer to output the action logits. The final action distribution of the policy network can be computed as:
\begin{equation}\label{policy}
    \pi(a_t^k | s_t^k) = \text{softmax}(f(e_t^{\text{node},k} \oplus e_t^{\text{ms},k})),
\end{equation}
where $(\cdot\oplus\cdot)$ denotes vector concatenation. This design allows the policy network to make context-aware decisions by integrating global scheduling history with local task-specific demands.

To enhance convergence and reduce the cost of inefficient exploration during early training, we adopt a two-phase learning strategy. The first phase pretrains the policy via behavior cloning (BC) using expert demonstrations, followed by online DRL in the second phase.

\subsection{Pretraining Policy Network}\label{sec:imitation} 
We train the policy network $\pi_\theta$ to imitate expert behavior using supervised learning. At each step $k$ in time slot $t$, the state $s_t^k = s_t^{\text{node},k} \cup s_t^{\text{ms},k}$ is processed by the enhanced architecture: a GRU models temporal dependencies within the node state $s_t^{\text{node},k}$, while a linear layer embeds the microservice-specific state $s_t^{\text{ms},k}$. The fused representation is used to predict a probability distribution over node selection. The imitation loss is defined as the negative log-likelihood of expert actions:
\begin{equation}\label{eq23}
    \mathcal{L}_{\text{IL}}(\theta) = - \sum\nolimits_{k\in\mathcal{K}_t} \log \pi_\theta\left(\pi_b(s_t^k)\mid s_t^k\right)
\end{equation}
where the $\pi_b(\cdot)$ is the expert policy. The training process is shown in Algorithm \ref{alg:bc_train}.

\begin{algorithm}[!t]
\caption{Behavior Cloning (BC) Pretraining}
\label{alg:bc_train}
\KwIn{Expert policy $\pi_b$}
\KwOut{Pretrained policy $\pi_\theta$}

Initialize policy parameters $\theta$\;
    \For{$t\in\{t_1,t_2,\cdots\}$}{
        Reset $h_t \leftarrow 0$, $\mathcal{L}_{\text{IL}}\leftarrow 0$\;
        \ForEach{microservice $k \in \mathcal{K}_t$}{
            Input $s_t^{node,k}$ to GRU, get $h_t$ by Eqn. (\ref{eq21})\;
            Predict $\pi_\theta(\pi_b(s^k_t) \mid s_t^k)$ by Eqn. (\ref{eq23})\;
            Compute loss $\mathcal{L}_{\text{IL}}+= -\log \pi_\theta(\pi_b(s^k_t) \mid s_t^k)$ \;
        }
        Update $\theta$ via gradient descent on $\mathcal{L}_{\text{IL}}$\;
    }
    \Return Policy $\pi_\theta$
\end{algorithm}

After behavior cloning, the learned parameters $\theta$ are used to initialize the actor network in the subsequent SAC training. GRU hidden states are reset at the beginning of each time slot to preserve intra-slot sequential structure while preventing inter-slot interference.

\subsection{Soft Actor-Critic (SAC)}
After behavior cloning, the pretrained policy is used to initialize the actor network in the online learning phase. We employ the SAC algorithm to further fine-tune the policy via interaction with the environment. SAC optimizes the expected long-term reward while encouraging sufficient exploration through entropy regularization.

The SAC algorithm operates within a Markov Decision Process (MDP) framework, where an agent interacts with an environment over discrete time steps. Each microservice $k$ at time slot $t$ involves observing state $s_{t}^{k}$, taking action $a_{t}^{k}$, receiving reward $r_{t}^{k}$, and transitioning to state $s_{t}^{k+1}$. SAC aims to maximize long-term rewards by adjusting policy entropy to balance exploration and exploitation. It uses a maximum entropy reinforcement learning framework to find the optimal policy $\pi$, mapping states to action distributions probabilistically to encourage exploration. The Bellman equation is crucial in MDP, relating soft action values to state value functions. Consequently, the relationship between the state $s_{t}^{k}$, action $a_{t}^{k}$, and time slot $t$ for the considered microservice $k$ can be described based on the Bellman equation:
\begin{equation}
    Q^{\pi } (s_{t}^{k},a_{t}^{k})=r^k_t+\gamma\cdot \mathbb{E}_{s_{t}^{k+1}\sim P(\cdot|s_t^k,a_t^k)}[V^{\pi}{(s_{t}^{k+1})}],
\end{equation}
and the soft value function as
\begin{equation}
    V^{\pi}({s_{t}^{k}})=\mathbb{E}_{a_{t}^{k}\sim \pi(\cdot|s^k_t)}[Q^{\pi}{(s_{t}^{k},a_{t}^{k})}-\beta\log{\pi(a_{t}^{k}|s_{t}^{k})]},
\end{equation}
where $P$ denotes the state transition probability, $\gamma$ is the discount factor in the equation, and $\beta$ is the entropy regularization coefficient. We consider $Q^\pi (s_{t}^{k},a_{t}^{k}) = Q_\omega (s_{t}^{k},a_{t}^{k})$ in the DNN where $\omega$ denotes the network parameters. The Q-function parameters are not constant, and the actor and critic are further updated according to the replay buffer regarding action and immediate reward in the replay buffer. Thus, the parameters can be trained by minimizing the loss function $J_Q(\omega)=\mathbb{E}_{(s_{t}^{k},a_{t}^{k},r_t^k,s^{k+1}_{t})\sim\mathbb{B}}$
\begin{equation}\label{critic}
\frac{1}{2}\left(Q_\omega(s_{t}^{k},a_{t}^{k})-(r^k_t+\gamma V(s^{k+1}_t))\right)^2,
\end{equation}
where $Q_\omega(s_{t}^{k},a_{t}^{k})$ denotes the soft Q-value, $\mathbb{B}$ denotes replay buffer, and $V(s^{k+1}_t)=\mathbb{E}_{a^{k+1}_t\sim\pi_{\theta}(\cdot|s^{k+1}_t)}$
\begin{equation}\label{eq26}
    \min_{j=1,2}Q_{\omega'_j}(s^{k+1}_t,a^{k+1}_t)-\beta\log\pi_\theta(a^{k+1}_t|s^{k+1}_t).
\end{equation}

In order to stabilize the iteration of the action value function, the Eqn. (\ref{eq26}) defines a target action-value function, in which $\omega'$ is obtained through an exponential moving average of $\omega$. The performance of DRL algorithms is heavily dependent on the policy. If the policy is optimal, the assigned microservice will be completed within the specified time, and the overall energy consumption will be reduced through reasonable microservice scheduling. Conversely, if the strategy is not optimal, microservice timeouts will become common, and overall energy consumption will be higher. Therefore, in order to improve the strategy, it is updated according to the Kullback-Leibler (KL) divergence:
\begin{equation}
    \pi(\cdot|s^k_t)=\arg\min_{\pi' \in \Pi }\text{KL}\left(\pi'(\cdot|s_{t}^{k})||\frac{\exp(Q^{\pi}(s_{t}^{k},\cdot)/\beta)}{Z^{\pi}(s_{t}^{k})}\right)
\end{equation}

\begin{algorithm}[!t]
    \textbf{Initialize:} $Q_{\omega_1}(s, a)$, $Q_{\omega_2}(s, a)$, $Q_{\omega_1'}$, and $Q_{\omega_2'}$ with weights $\omega_1' = \omega_1$ and $\omega_2' = \omega_2$\;
    \textbf{Initialize:} policy $\pi_{\theta}(a|s)$ with weight $\theta$\;
    \For{each epoch}{
        Retrieve current state $s^{node}_1$\;
        \For{$t\in\{t_1,t_2,\cdots\}$}{
            \textbf{Initialize:} hidden state of GRU $h_t$\;
            \For{$k\in\{k_1, k_2, \cdots, k_{K_t}\}$}{
                Obtain state of node $s_{t}^{node,k}$ form environment\;
                $s_t^k=s_t^{node,k}\cup s_{t}^{ms,k}$\;
                Obtain $\pi(\cdot|s^k_t)$ based on Eqn. (\ref{policy})\;
                Get offloading action $a_{t}^{k}$ by Algorithm \ref{a2}\;
            }
            Execute $a_t=\{a^{k_1}_t,a^{k_2}_t,\cdots,a^{k_{K_t}}_t\}$\;
            \For{$k\in\{k_1, k_2, \cdots, k_{K_t}\}$}{
                Get reward $r_{t}^{k}$\;
                Save $(s_{t}^{k}, a_{t}^{k}, r_{t}^{k}, s_{t}^{k+1})$ in replay memory $\mathbb{B}$\;
            }
        }
        Update $J_Q(\omega)$ and $\theta$ by using $\theta \leftarrow \theta + \eta_a\cdot\nabla{_\theta} J_\pi(\theta)$ and $\omega \leftarrow \omega + \eta_c\cdot\nabla{_\omega} J_Q(\omega)$\;
        Update soft action value function $\omega'$\;
    }
    \caption{SAC-Based Microservices Offloading}
    \label{a1}
\end{algorithm}

In the provided text, $\Pi$ represents a collection of policies that correspond to Gaussian parameters, and the $\text{KL}$ term quantifies the information loss during the approximation process. The KL divergence can be reduced further by adjusting the policy parameters through the following policy parameter updates: $J_{\pi}(\theta)=$
\begin{equation}\label{actor}
    \mathbb{E}_{s_{t}^{k}\sim\mathbb{B}}[\mathbb{E}_{a_{t}^{k}\sim \pi_{\theta}(\cdot|s^k_t)}[\beta \log_{}{(\pi_{\theta }(a_{t}^{k}|s_{t}^{k}))-Q_{\omega}(s_{t}^{k},a_{t}^{k})]]}. 
\end{equation}
The policy iteration is continued until it reaches the optimal value and converges with the maximum entropy. The interactions between the environment, the value function, and the policy network are shown in Fig. \ref{fig1}. The detailed SAC-based microservices offloading algorithm is given in Algorithm \ref{a1}, which is different from the original SAC algorithm in action execution. As shown in line 12 of Algorithm \ref{a1}, we can observe the reward of each microservice after executing all actions in this time slot. This is because the inter-dependency among tasks in the same time slot.


\begin{algorithm}[!t]
    Get action distribution $\pi(\cdot|s^k_t)$ from actor network\;
    \For{$n \in\mathcal{N}$}{
        Calculate $u_{n}=\kappa^{c}_{n}\cdot{\kappa^{m}_{n}}\cdot{\kappa^{i}_{n}}$\; 
        and $u_t^k[n]\leftarrow u_{n}$\;
    }
    $\pi'(\cdot|s^k_t) \leftarrow \pi(\cdot|s^k_t) \odot u_t^k$ and normalize to get $\pi'(\cdot|s^k_t)$\;
    Sample $a_t^k$ from $\pi'(\cdot|s^k_t)$\;
    \Return $a_t^k$\;
    \caption{Action Selection}
    \label{a2}
\end{algorithm}

\subsection{Action Selection}
When selecting actions, the agent samples according to the probability output of the policy network, and it does not judge whether the action is reasonable. Some constraints should be added to the action selection process to avoid some of those unacceptable actions. This problem was solved by defining an Action Mask $u_t^k$ to fit the constraints. The mask is a one-dimensional Boolean vector with the same length as the action space $A$. This Action Mack is updated before each action selection; its value depends on the feasibility of a specific action in the whole action space. An example of an action mask $u_t^k=[1,0,1,0,0,1,\cdots, u_{N-1},u_{N}]$. By employing a mask to set a portion of the action probabilities generated by the policy network to zero, we ensure that these nodes will not be chosen as actions. The calculation of each Boolean value $u_{n}$ in the Action Mask $u_t^k$ will be shown in the following.

First, for the node $n$, if there are microservices already processing on it and it has no idle CPU frequency, then the node is considered a busy node, which can be obtained as $\kappa^{c}_{n}=\mathbb{I}{\{F_n(t)>0\}}$, where $\mathbb{I}(\cdot)$ is an indicator function and $u^{c}_{n}=1$ means that the node is acceptable, otherwise it could not be selected as an action. Second, when the node is experiencing a shortage of memory and is not available for the microservices $k$, which can be expressed as $\kappa^{m}_{n}=\mathbb{I}{\{M^k_n(t)-m_k>0\}}$. The storage considerations for images are nearly the same, but it's crucial to assess whether the node already contains the microservices $k$ required image $i_k$. Thus, we have $\kappa^{i}_{n}=\mathbb{I}\{\{D^k_n(t) - s_{i_k}\geq 0\}\vee\{x_{n,i}==0\}\}$.
If $\kappa^{c}_{n}$, $\kappa^{m}_{n}$, and $\kappa^{i}_{n}$ are all equal to 1, the action is acceptable. Otherwise, it is not a good action. To sum up, each Boolean value $u_{n}$ in $u_t^k$ can be summarized as $u_{n}=\kappa^{c}_{n}\cdot{\kappa^{m}_{n}}\cdot{\kappa^{i}_{n}}$. The action selection is shown in Algorithm \ref{a2}.

\section{Performance Evaluation}\label{sec5}
In this section, we evaluate the performance of the proposed algorithm through extensive experiments under various settings. We compare our method with several baselines in terms of convergence speed, total completion time, energy consumption, image download latency, and on-time completion ratio.

\subsection{Experimental Settings}

\subsubsection{Parameter Settings}
All IoT devices are heterogeneous and randomly distributed, and the default number of nodes is $15$. The node's CPU frequency is set between $[3,6.5]$ GHz, and memory is set between $[80,180]$ GB. The network bandwidth is limited to between $[2, 6]$ Gbps. Each image requires a maximum of $10$ MB, the maximum microservice size is set to $100$ MB, and the maximum transfer power is set to $0.0001995$.

\begin{table}[h]
    \caption{Hyperparameters}
    \label{table2}
    \centering
    \begin{tabular}{ll}
        \toprule
        \textbf{Hyperparameter} & \textbf{Value} \\
        \midrule
        Replay Buffer Size & 22000 \\
        Minimal Update Size & 1500 \\
        Batch Size & 3000 \\
        Actor GRU Dim & 64 \\
        Actor FC Layers & (128, 64, 32) \\
        Critic FC Layers & (64, 16) \\
        Actor LR & $1\!\times\!10^{-5}$ \\
        Critic LR & $3\!\times\!10^{-4}$ \\
        Temp LR ($\alpha$) & $1\!\times\!10^{-4}$ \\
        Optimizer & Adam \\
        Discount ($\gamma$) & 0.98 \\
        Soft Update ($\tau$) & 0.005 \\
        Init $\alpha$ & $\log(0.01)$ \\
        Target Entropy & $-1$ \\
        \bottomrule
    \end{tabular}
\end{table}

\subsubsection{Baselines}
To compare performance, several baselines were used. \textit{The word ``Hybrid'' indicates adopting our two-way enhanced embedding design in the policy network, and ``BC'' indicates adopting IL-based Pretrain in the implementation.} The details are as follows.
\begin{itemize}
    \item \textbf{Hybrid\_SAC:} This is our proposed GRU-based SAC algorithm, as shown in Algorithm~\ref{a1}. It combines GRU-based encoding of node states with task embedding and fully connected layers. Hyperparameters are listed in Table~\ref{table2}.
     \item \textbf{BC\_Hybrid\_SAC:} A two-stage implementation. The model shares the same architecture as Hybrid\_SAC, and is first trained via the BC method proposed in this paper, before being trained online with RL.
    \item \textbf{Hybrid\_PPO:} A PPO-based algorithm that shares the same GRU-based actor architecture as Hybrid\_SAC.
     \item \textbf{BC\_Hybrid\_PPO:} A two-stage implementation. The model is identical to Hyper\_PPO and is first trained using BC before applying PPO updates.
    \item \textbf{GRU\_SAC:} A variant of SAC in which the actor network is fully composed of GRU layers without the hybrid task-node structure described in this paper. The critic network remains fully connected. All hyperparameters are the same as in Table~\ref{table2}.
    \item \textbf{BC\_GRU\_SAC:} A two-stage implementation. The model is structurally the same as GRU\_SAC and is pre-trained with BC before SAC training.
    \item \textbf{SAC:} A standard SAC algorithm with both actor and critic implemented using fully connected layers. The actor network structure follows $(128, 64, 32)$.
    \item \textbf{BC\_SAC:} A two-stage implementation. The model uses the same fully connected architecture as SAC and is first trained using BC, followed by SAC-based RL.
    \item \textbf{PPO \cite{schulman2017proximal}:} The policy and value networks are implemented using two fully connected layers with sizes $(128, 64, 32)$.
     \item \textbf{BC\_PPO:} A two-stage implementation. The model has the same structure as PPO and is trained via BC before PPO optimization.
    \item \textbf{DQN \cite{mnih2015human}:} The Q-network consists of two fully connected layers with sizes $(128, 64, 32)$. BC is not applicable to this value-based method.
    \item \textbf{Greedy:} Uses the expert policy described in Algorithm~\ref{alg:expert_policy} to make decisions. At each step, it selects node $n$ that minimizes the image download waiting time for microservice $k$, while satisfying resource constraints.
\end{itemize}

\subsubsection{Offline Data Collection}
We design a rule-based expert policy $\pi_b$ that selects the most suitable edge node for each microservice $k$ by jointly considering delay tolerance and energy efficiency. The scoring function $\mathcal{S}_n$ evaluates each node $n$ based on estimated execution delay and energy cost. To ensure that expert decisions satisfy system constraints, we use the action feasibility mask defined in Algorithm~\ref{a2}. This unifies the action constraints across both IL and RL stages.

During offline data collection, the expert estimates the execution time $T_{n,k}^{\text{comp}}$ by assuming that the CPU frequency $F_n(t)$ of node $n$ will be equally divided among all tasks that currently have been assigned to it. This means the estimated computation time for task $k$ is based on the number of tasks at the time of scheduling, ignoring the fact that more tasks may be assigned to the same node later in the same time slot. This assumption introduces a bias—later tasks will reduce the actual CPU share for earlier tasks—but provides a practical approximation that enables efficient data collection and guides the imitation learning process. The expert decision logic is described in Algorithm~\ref{alg:expert_policy}.

\begin{algorithm}[!t]
\caption{Expert Policy $\pi_b$}
\label{alg:expert_policy}
\KwIn{State $s_t^k$, microservice $k$}
\KwOut{Expert action $a_t^{k,\text{expert}}$}

Initialize best score $\mathcal{S}^* \leftarrow -\infty$, best node $n^* \leftarrow -1$ \;

\ForEach{node $n \in \mathcal{N}$}{
    \If{$u_n = 0$ by Algorithm~\ref{a2}}{\textbf{continue}}

    Compute delay margin $d_n = l_k - (T_{n,k}^{\text{comm}} + T_{n,k}^{\text{down}} + T_{n,k}^{\text{comp}})$\;
    Compute energy cost $e_n$\;
    Compute score $\mathcal{S}_n = \alpha d_n - e_n$\;

    \If{$\mathcal{S}_n > \mathcal{S}^*$}{
        $\mathcal{S}^* \leftarrow \mathcal{S}_n$\;
        $n^* \leftarrow n$\;
    }
}
\Return $a_t^{k,\text{expert}} \leftarrow n^*$
\end{algorithm}

\subsection{Experimental results: Two-Phase Learning}
\begin{figure}[!t]
    \centering
    \includegraphics[width=\linewidth]{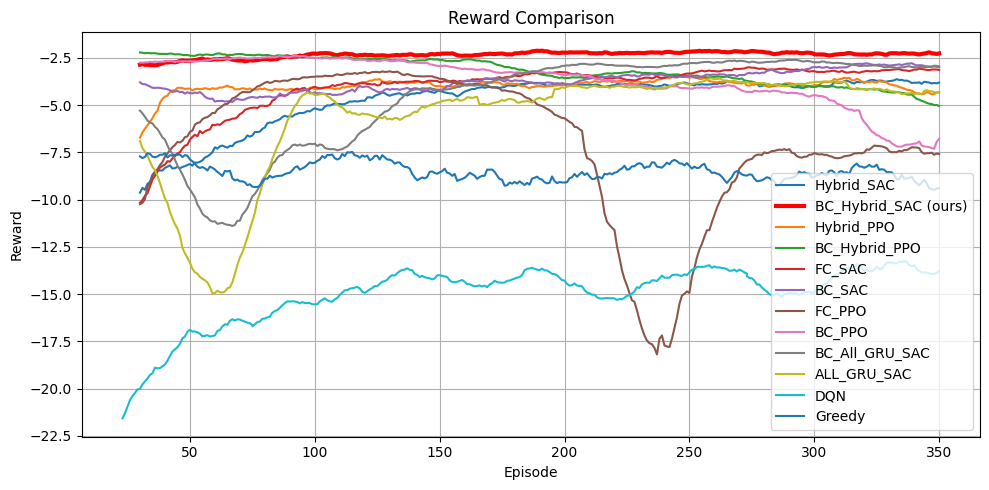}
    \caption{Convergence performance of different scheduling algorithms over 350 episodes. The plot shows smoothed reward trajectories under identical environment settings. Compared models include our GRU-based Hybrid\_SAC and its behavior cloning variant (BC\_Hybrid\_SAC), GRU-based PPO (Hybrid\_PPO), standard FC-based SAC and PPO, and their BC variants. GRU\_SAC removes the hybrid structure to assess its impact. DQN and the rule-based Greedy strategy serve as value-based and heuristic baselines. BC\_Hybrid\_SAC achieves the highest reward and fastest convergence.}
    \label{rewardcomp}
\end{figure}

We present the convergence results for all baseline methods in Fig.~\ref{rewardcomp}, under the setting of 15 nodes, 5--20 microservices per time slot, and $\alpha=1$. Among all methods, BC\_Hybrid\_SAC achieves the highest and most stable reward, demonstrating the effectiveness of combining a GRU-based architecture with behavior cloning. Hybrid\_SAC also converges quickly, stabilizing after around 40 episodes. While Hybrid\_PPO shows similarly fast convergence, its average reward is slightly lower. In contrast, fully connected SAC and PPO converge more slowly, and PPO exhibits overfitting behavior after 160 episodes, resulting in performance degradation. DQN, being value-based, converges slowly and yields the lowest reward overall. The rule-based Greedy strategy maintains a constant reward across episodes. Overall, GRU-based models with behavior cloning (especially BC\_Hybrid\_SAC) significantly outperform other approaches in both convergence speed and final reward.

\begin{figure}[!t]
    \centering
    \subfigure[$\#$Nodes = 12]{
        \includegraphics[width=0.46\linewidth]{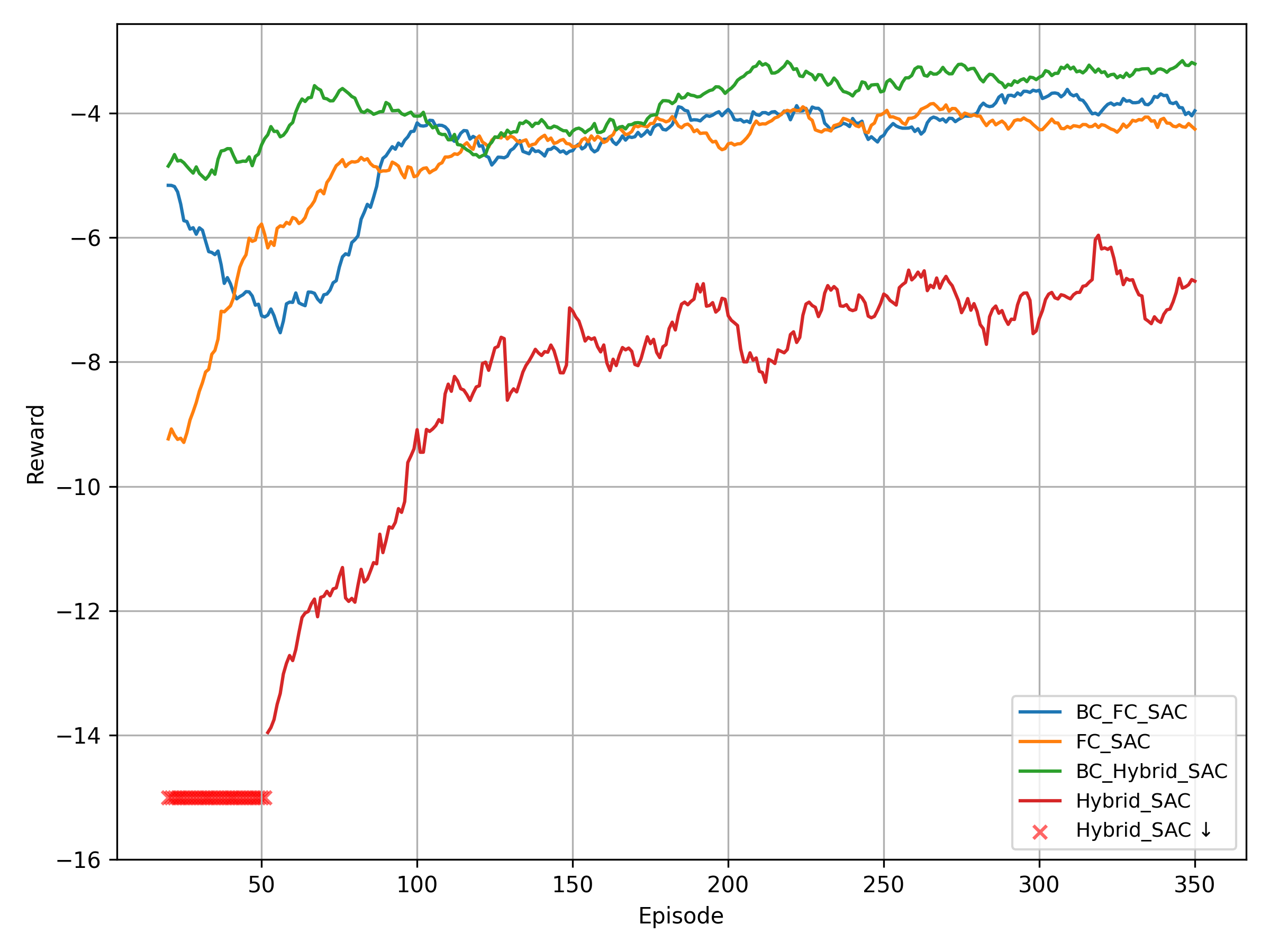}
        \label{Compare_BC_node12}
    }
    \subfigure[$\#$Nodes = 15]{
        \includegraphics[width=0.46\linewidth]{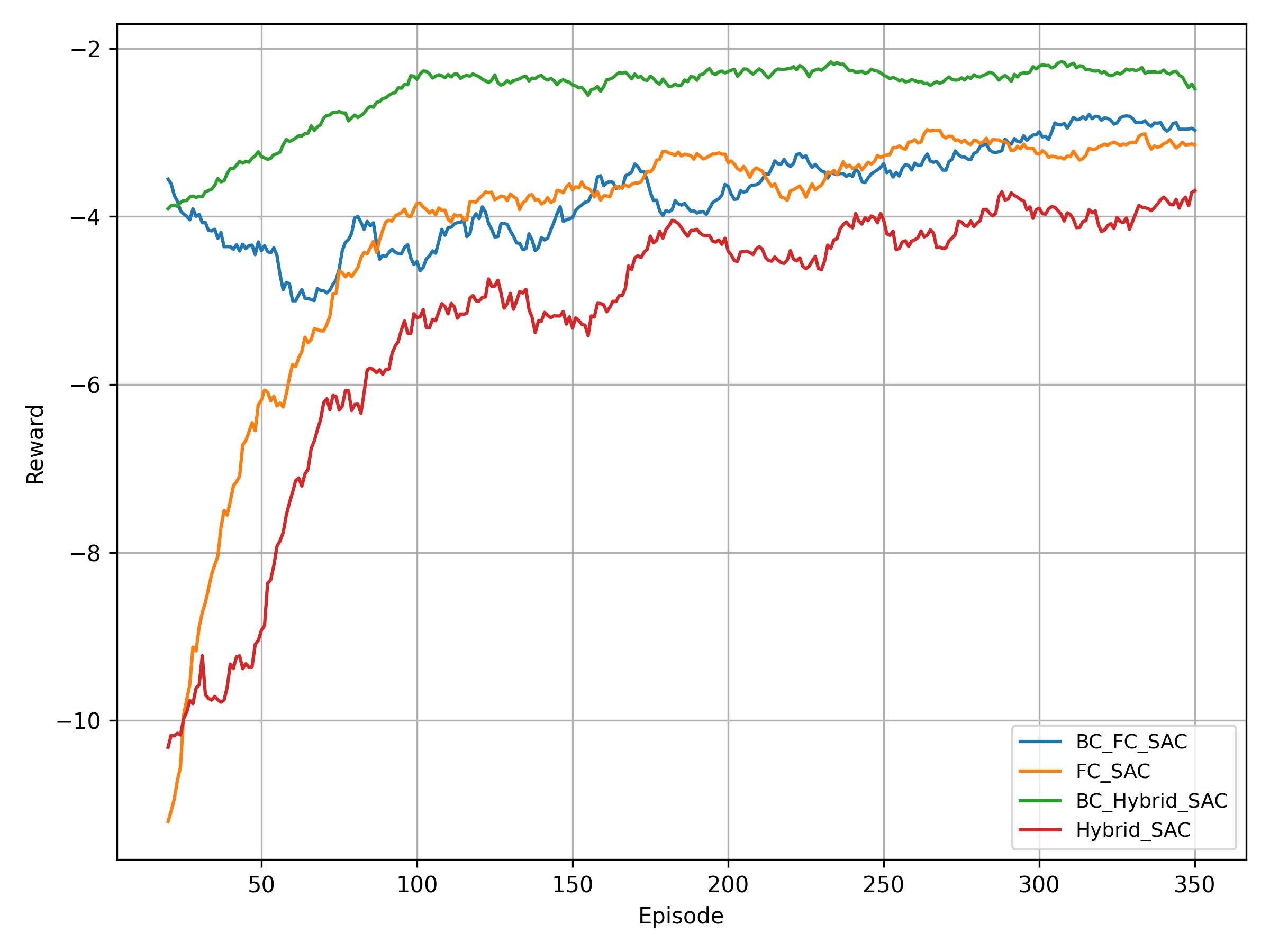}
        \label{Compare_BC_node15}
    }

    \subfigure[$\#$Nodes = 18]{
        \includegraphics[width=0.46\linewidth]{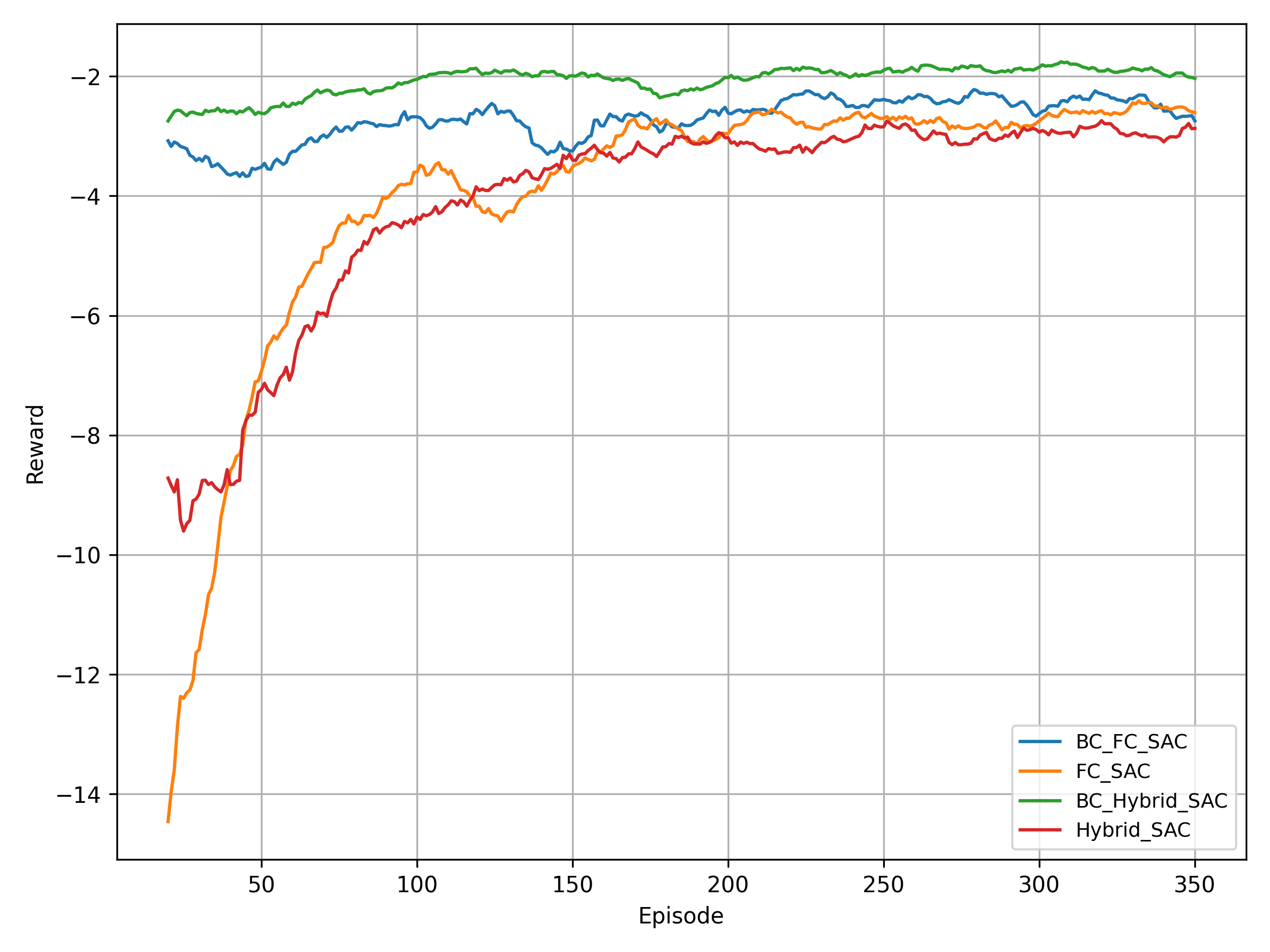}
        \label{Compare_BC_node18}
    }
    \subfigure[$\#$Nodes = 20]{
        \includegraphics[width=0.46\linewidth]{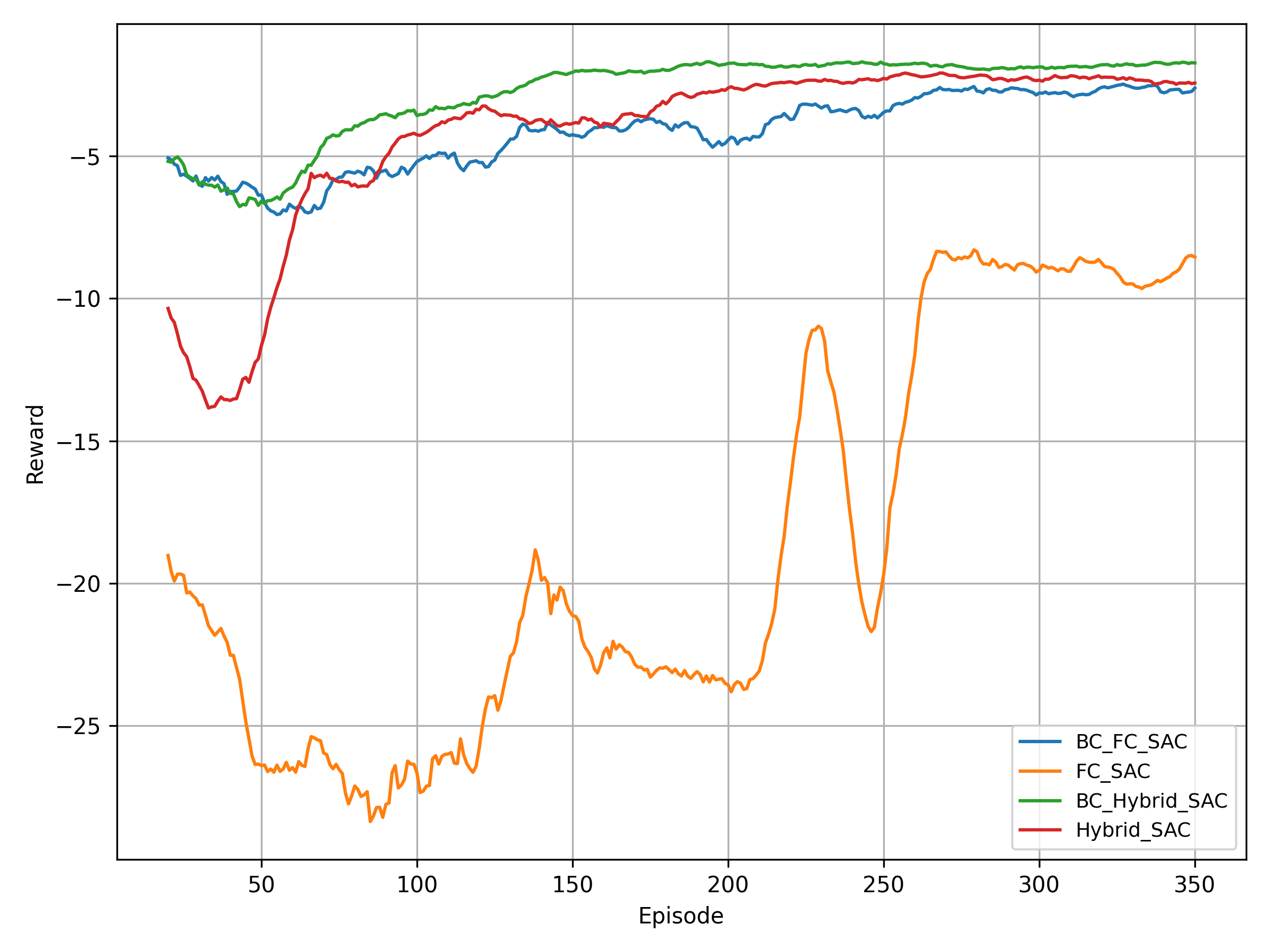}
        \label{Compare_BC_node20}
    }
    \caption{Comparison of convergence performance across different node settings with and without Behavior Cloning. Each subfigure shows the smoothed reward trajectories of Hybrid\_SAC and SAC models, both with and without BC initialization.}

    \label{BCrewardcomp}
\end{figure}

\subsubsection{Effectiveness of Behavior Cloning}

To further evaluate the impact of offline imitation learning on training efficiency, we compare SAC-based models with and without behavior cloning (BC) under different policy architectures. Specifically, we assess the performance of Hybrid\_SAC and SAC, each trained either from scratch or with BC initialization. During the BC phase, expert demonstrations were collected over 10 offline episodes, generating approximately 20{,}000 state-action pairs. Each model was pre-trained for 20 epochs using the Adam optimizer (learning rate $1 \times 10^{-4}$, batch size 64). The results are shown in Fig.~\ref{BCrewardcomp}, where the $x$-axis denotes training episodes and the $y$-axis shows the smoothed reward trajectories under different initialization strategies.

As illustrated, models initialized with BC consistently converge faster and reach higher rewards across all node configurations. In the 15-node scenario (Fig.~\ref{Compare_BC_node15}), BC\_Hybrid\_SAC achieves a final reward of approximately $-3.2$ by episode 350, whereas Hybrid\_SAC trained from scratch only reaches around $-4.2$. The benefit becomes more prominent in higher-dimensional cases: in the 20-node setting (Fig.~\ref{Compare_BC_node20}), BC\_Hybrid\_SAC quickly surpasses a reward of $-4.5$ within the first 100 episodes and eventually stabilizes near $-2.5$. In contrast, the non-BC variant shows severe fluctuations and ends below $-8.5$.

Similar trends are observed under the 12-node and 18-node settings (Fig.~\ref{Compare_BC_node12} and Fig.~\ref{Compare_BC_node18}). For example, in the 12-node case, BC\_Hybrid\_SAC converges to around $-3.5$, while Hybrid\_SAC without BC remains unstable around $-7.8$, with frequent performance drops. The performance gap is especially evident in smaller node settings, where suboptimal early actions by a randomly initialized policy lead to delayed learning and reduced stability. In contrast, BC initialization provides a strong prior that guides early exploration and accelerates convergence.
Overall, these results demonstrate that behavior cloning significantly improves both learning efficiency and final performance. When integrated into a two-stage training framework, BC offers a robust initialization that enhances sample efficiency and policy stability. These advantages make the proposed Hybrid\_SAC framework well-suited for dynamic and heterogeneous edge environments, where microservice scheduling requires high-dimensional, real-time decision making.

\begin{figure}[!t]
    \centering
    \subfigure[$\#$Nodes = 12]{
        \includegraphics[width=0.46\linewidth]{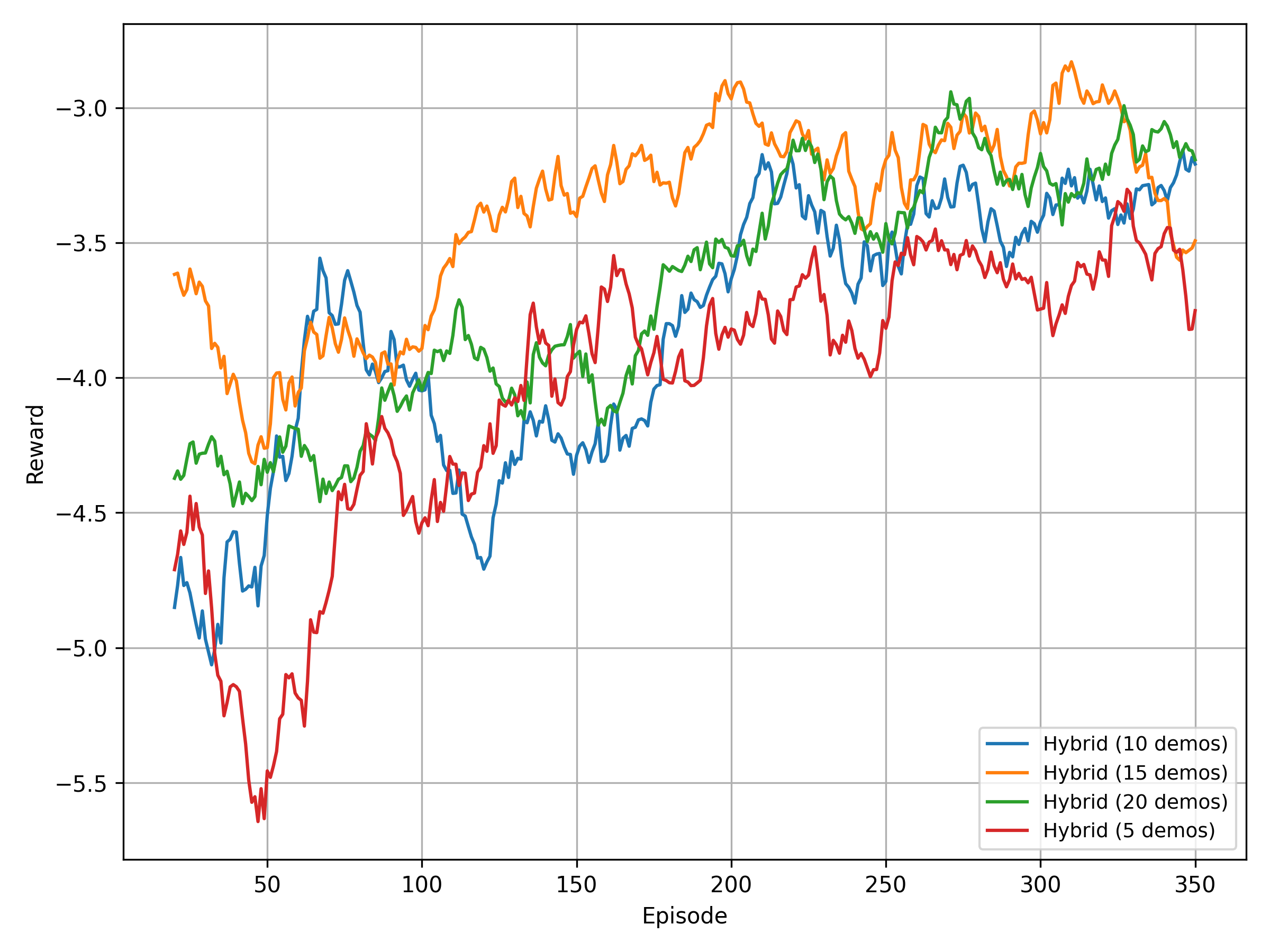}
        \label{Compare_dome_size_node12}
    }
    \subfigure[$\#$Nodes = 15]{
        \includegraphics[width=0.46\linewidth]{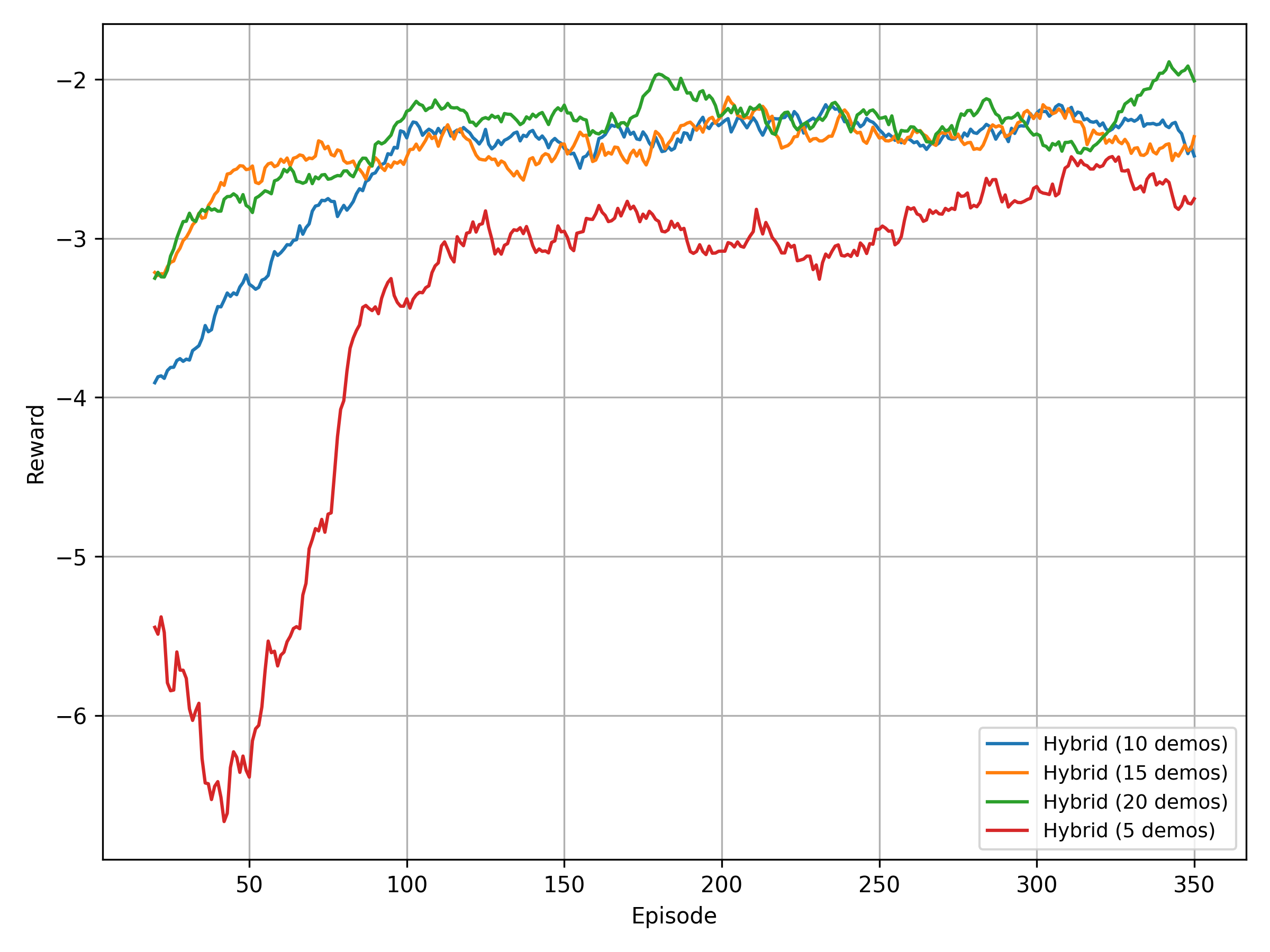}
        \label{Compare_dome_size_node15}
    }

    \subfigure[$\#$Nodes = 18]{
        \includegraphics[width=0.46\linewidth]{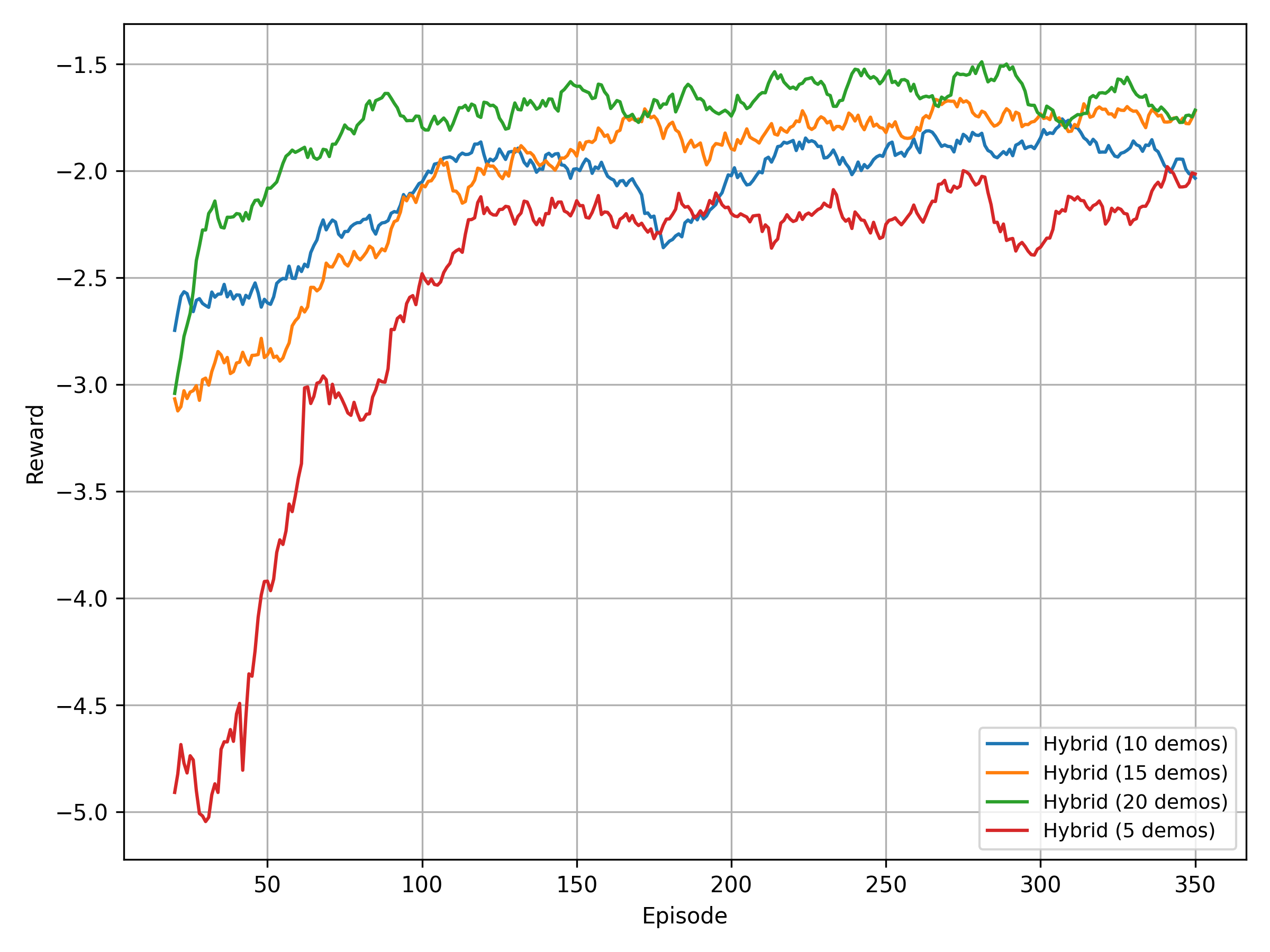}
        \label{Compare_dome_size_node18}
    }
    \hfill
    \subfigure[$\#$Nodes = 20]{
        \includegraphics[width=0.46\linewidth]{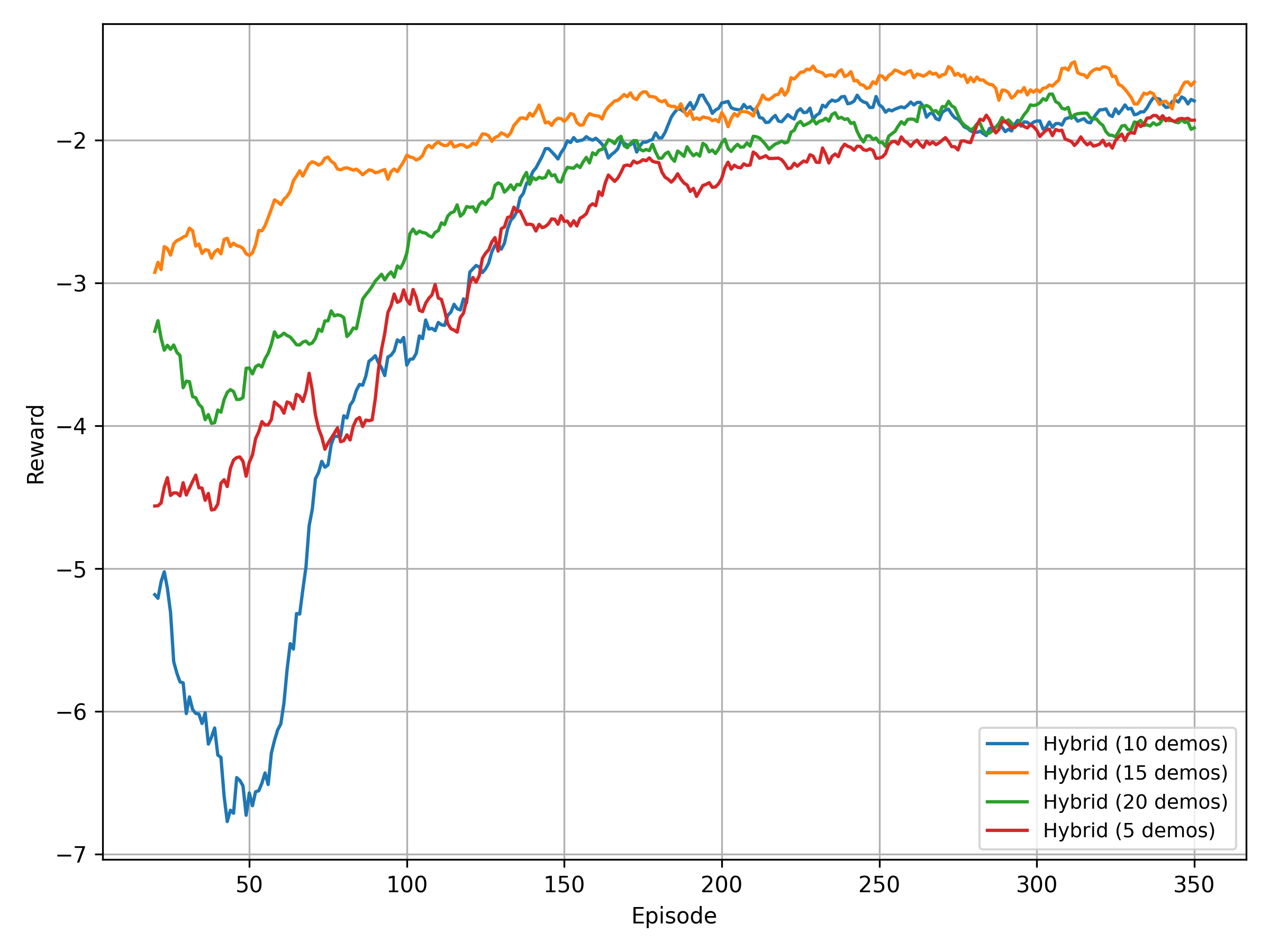}
        \label{Compare_dome_size_node20}
    }
    \caption{Convergence performance comparison under different node numbers with varying sizes of expert demonstration data. Each subfigure illustrates the reward trajectories of Hybrid\_SAC models pretrained with 5, 10, 15, and 20 demonstration episodes, highlighting the impact of demonstration quantity on offline behavior cloning effectiveness.}
    \label{Compare_dome_size_nodes}
\end{figure}

\subsubsection{Impact of Expert Demonstration Count}

To examine how the number of expert demonstrations affects policy pretraining quality, we conducted additional experiments under varying node settings. Specifically, we collected 5, 10, 15, and 20 expert episodes for behavior cloning, where each episode contains approximately 2{,}000 state-action pairs. The resulting policies were used to initialize BC\_Hybrid\_SAC and subsequently fine-tuned via reinforcement learning.

The convergence results are shown in Fig.~\ref{Compare_dome_size_nodes}. Across all configurations, BC\_Hybrid\_SAC initialized with more demonstrations converges faster and achieves higher final rewards. In the 15-node setting (Fig.~\ref{Compare_dome_size_node15}), the model trained with 20 expert episodes reaches a final reward of approximately $-2.0$, while the one trained with only 5 episodes converges to around $-3.5$. This trend becomes even more pronounced as the number of nodes increases. For example, in the 20-node case (Fig.~\ref{Compare_dome_size_node20}), the reward gap between the 5-demo and 20-demo variants exceeds 1.5, highlighting the importance of strong initialization in high-dimensional scheduling scenarios.

Nonetheless, we observe diminishing returns when increasing demonstrations from 15 to 20. This saturation suggests that a moderate number of high-quality expert episodes (around 15) may already suffice for effective policy initialization, avoiding excessive offline data collection costs while maintaining training efficiency.

\begin{figure}[!t]
    \centering
    \subfigure[$\#$Nodes = 12]{
        \includegraphics[width=0.46\linewidth]{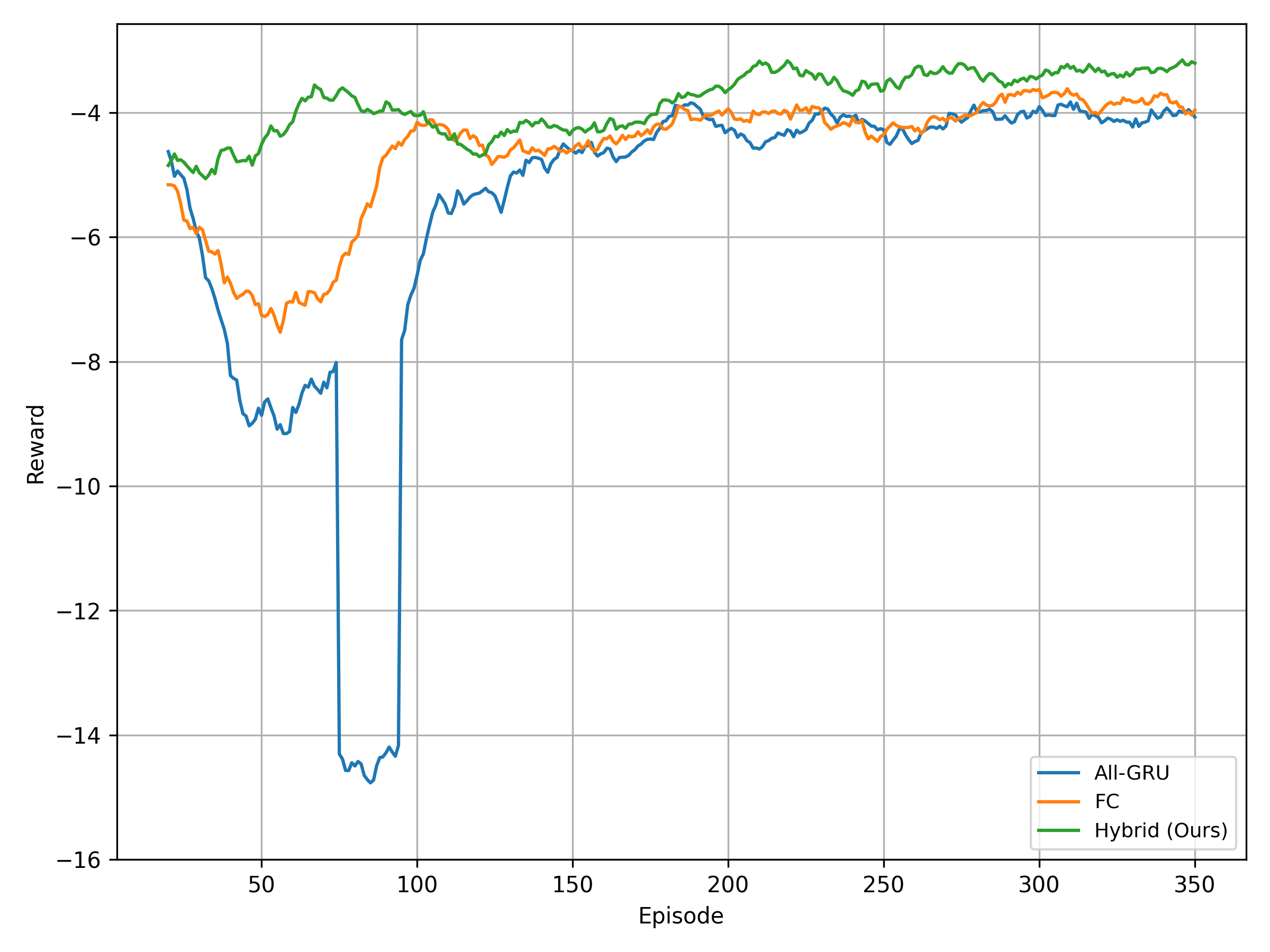}
        \label{Compare_Structure_node12}
    }
    \subfigure[$\#$Nodes = 15]{
        \includegraphics[width=0.46\linewidth]{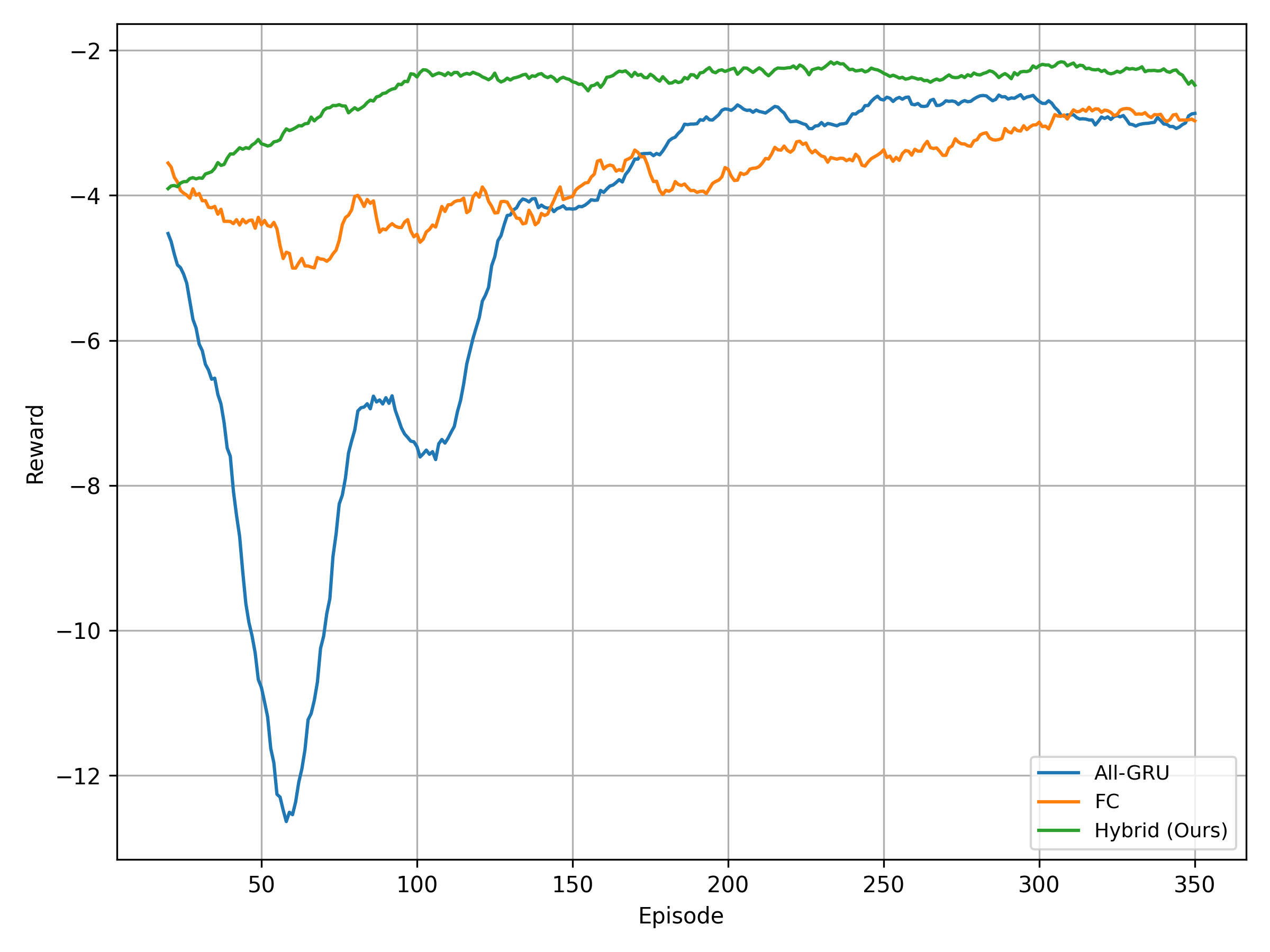}
        \label{Compare_Structure_node15}
    }

    \subfigure[$\#$Nodes = 18]{
        \includegraphics[width=0.46\linewidth]{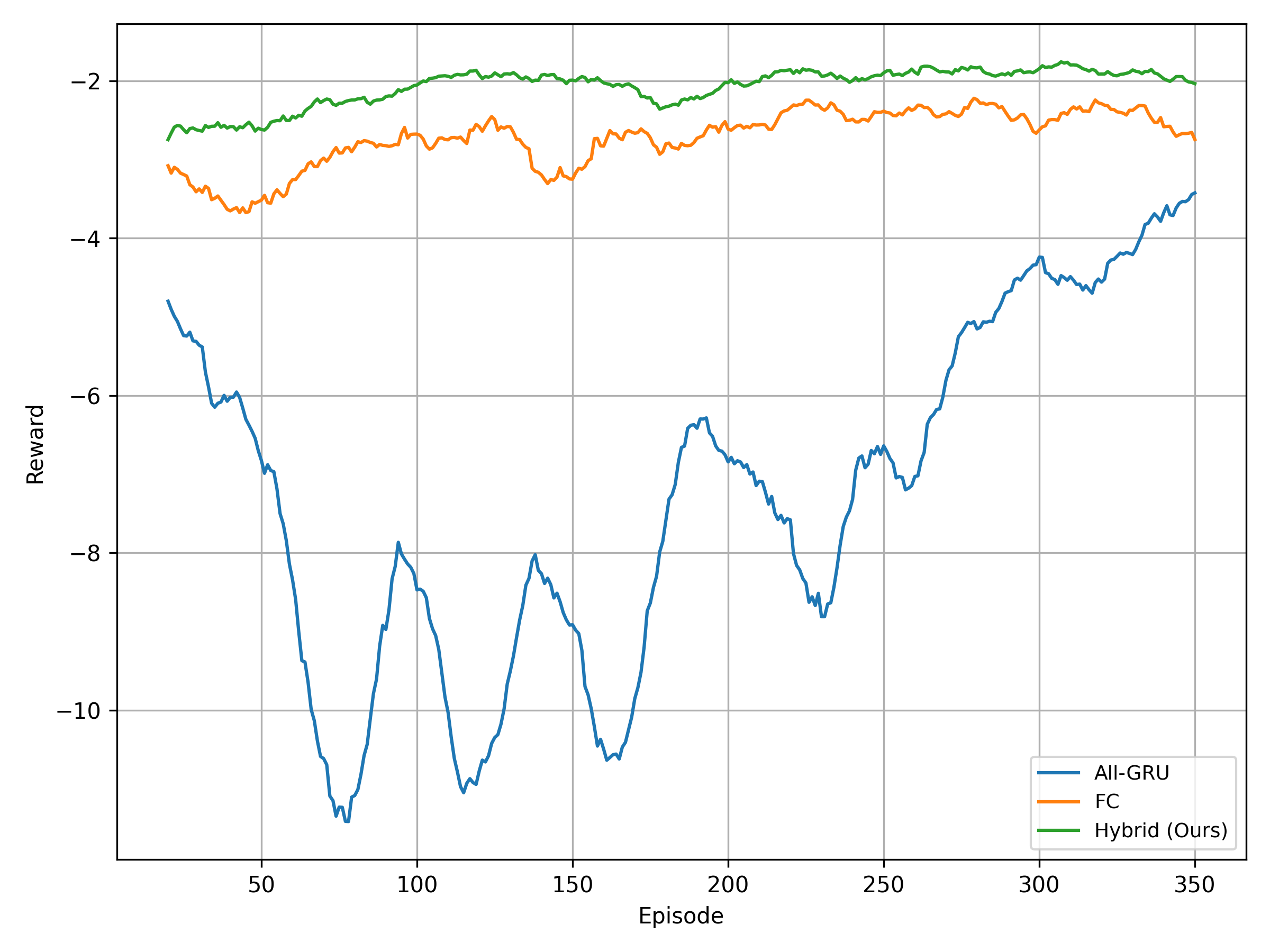}
        \label{Compare_Structure_node18}
    }
    \subfigure[$\#$Nodes = 20]{
        \includegraphics[width=0.46\linewidth]{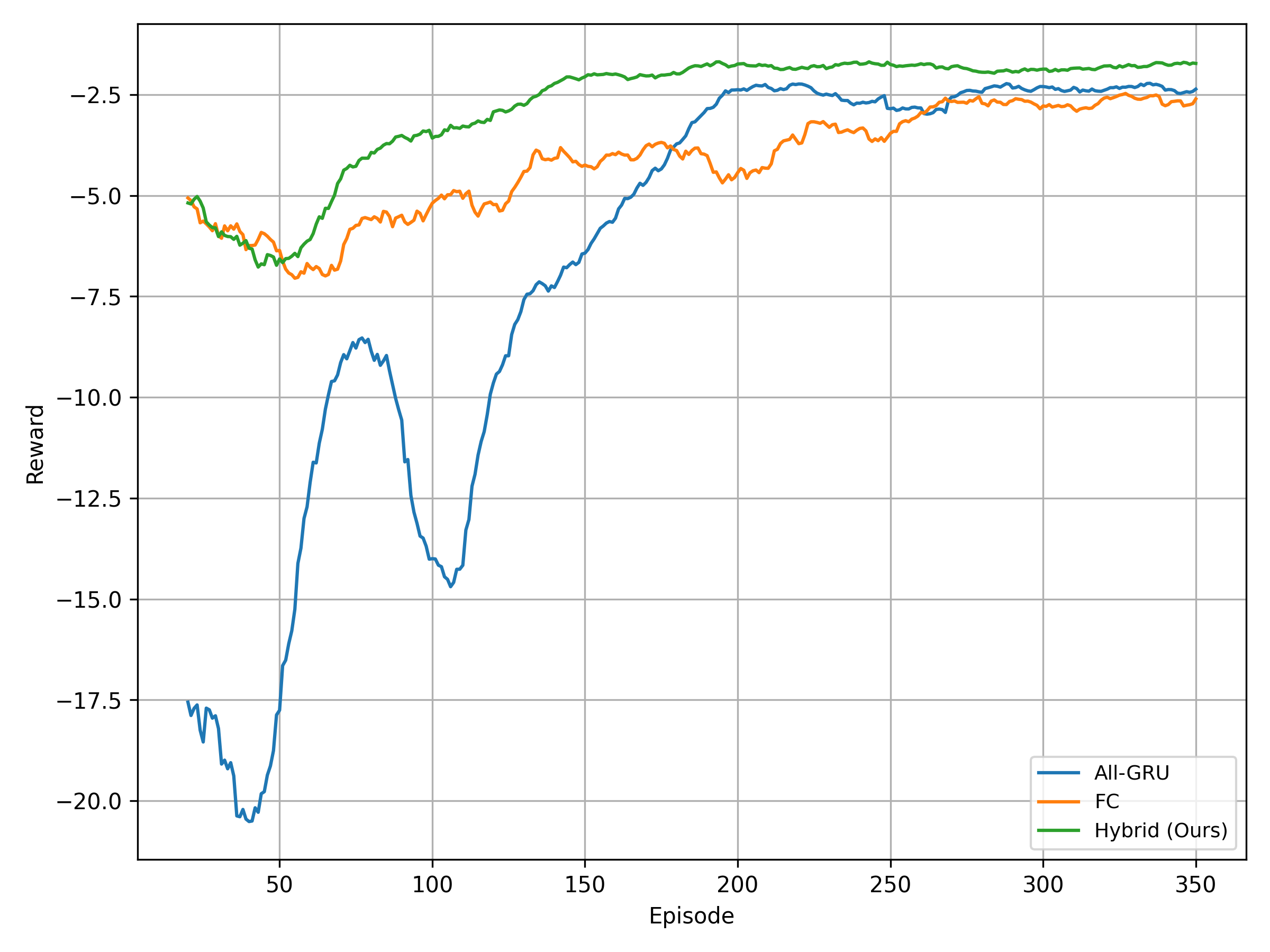}
        \label{Compare_structure_node20}
    }
    \caption{Comparison of convergence performance across different node settings for BC-initialized policy network structures. Each subfigure shows the smoothed reward trajectories of BC\_Hybrid\_SAC, BC\_GRU\_SAC, and BC\_SAC models.}
    \label{Compare_Structure_nodes}
\end{figure}

\subsubsection{Ablation Study on Policy Network Structure}

To evaluate the effectiveness of our proposed enhanced policy network design, we conducted an ablation study comparing three behavior cloning-initialized variants: (1) a fully connected policy network (BC\_SAC), where all input features are processed through linear layers; (2) a GRU-only policy network (BC\_GRU\_SAC), where all features are encoded sequentially using a GRU module; and (3) the proposed enhanced design (BC\_Hybrid\_SAC), which encodes slow-varying node states via GRU and fast-changing microservice features through linear embeddings. All models were pretrained with behavior cloning using 10 expert episodes and subsequently fine-tuned via SAC.

The convergence results under different node settings are shown in Fig.~\ref{Compare_Structure_nodes}. Across all configurations, BC\_Hybrid\_SAC consistently achieves faster convergence and higher final rewards than the other two architectures. The BC\_GRU\_SAC variant suffers from unstable training and large fluctuations, likely due to the difficulty of modeling fast-changing task inputs with a sequential encoder. BC\_SAC shows more stable learning but converges to lower reward levels, indicating that disregarding temporal dependencies in node states results in the loss of valuable scheduling information. In contrast, BC\_Hybrid\_SAC effectively balances temporal modeling and responsiveness to dynamic task states, enabling more efficient and robust learning.

Moreover, the performance gap between architectures becomes increasingly pronounced as the number of nodes grows. Larger node settings lead to higher-dimensional action spaces, increasing the complexity of decision making. Under small action spaces (e.g., 12 nodes), BC\_SAC and BC\_Hybrid\_SAC both show relatively stable learning, though BC\_Hybrid\_SAC still achieves better final rewards. As the node count increases to 18 or 20, BC\_GRU\_SAC exhibits significant instability, while BC\_SAC also struggles to maintain high performance. In contrast, BC\_Hybrid\_SAC remains stable and continues to outperform other models, demonstrating superior scalability in complex scheduling environments.
These findings validate the design rationale of BC\_Hybrid\_SAC: selectively applying sequential modeling to slow-changing node states, while processing fast-varying microservice inputs independently, enables the policy to generalize more effectively across dynamic, high-dimensional action spaces.

\begin{figure}[!t]
    \centering
    \subfigure[$\#$Nodes = 12]{
        \includegraphics[width=0.46\linewidth]{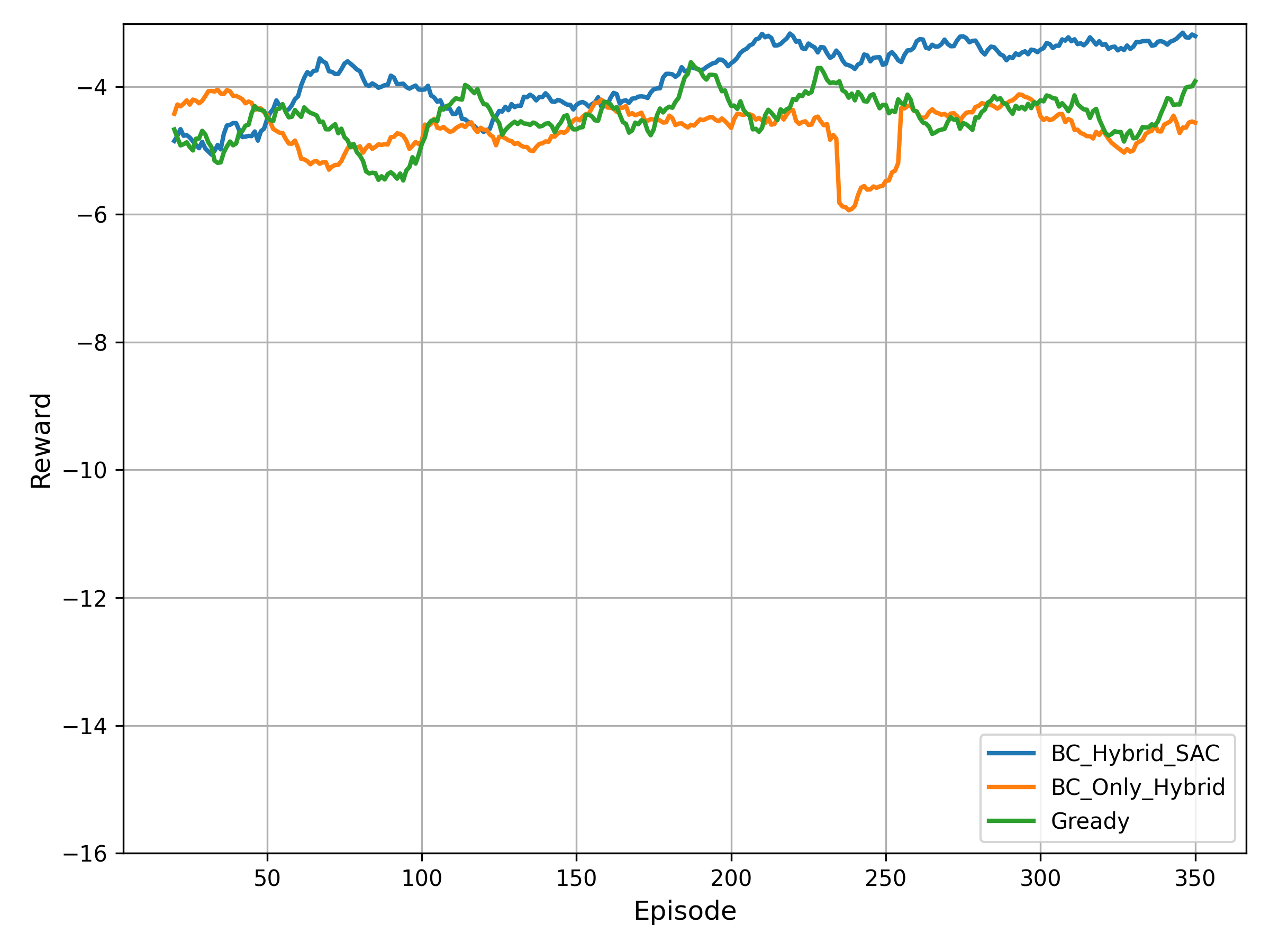}
        \label{Compare_RL_node12}
    }
    \subfigure[$\#$Nodes = 15]{
        \includegraphics[width=0.46\linewidth]{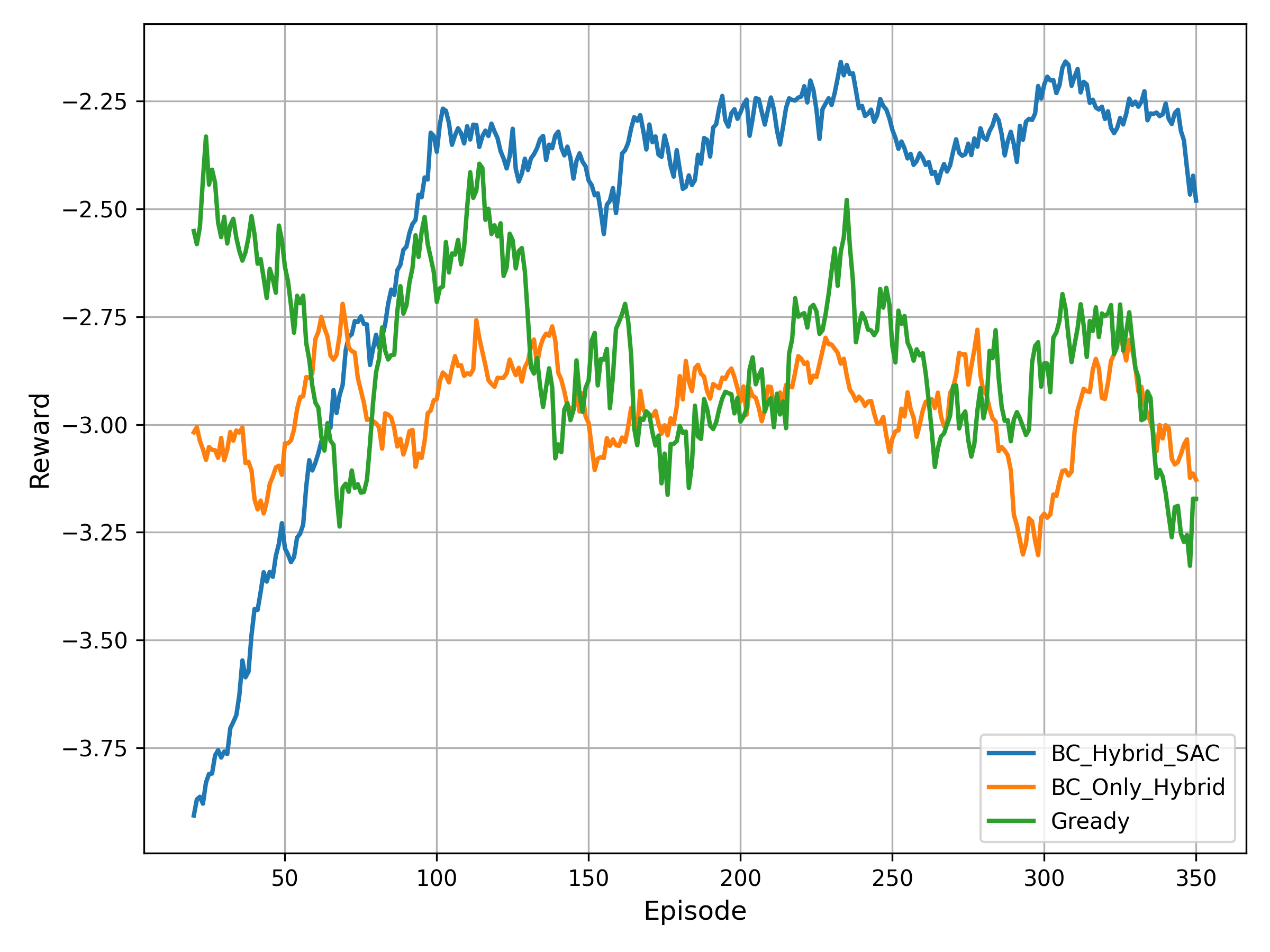}
        \label{Compare_RL_node15}
    }

    \subfigure[$\#$Nodes = 18]{
        \includegraphics[width=0.46\linewidth]{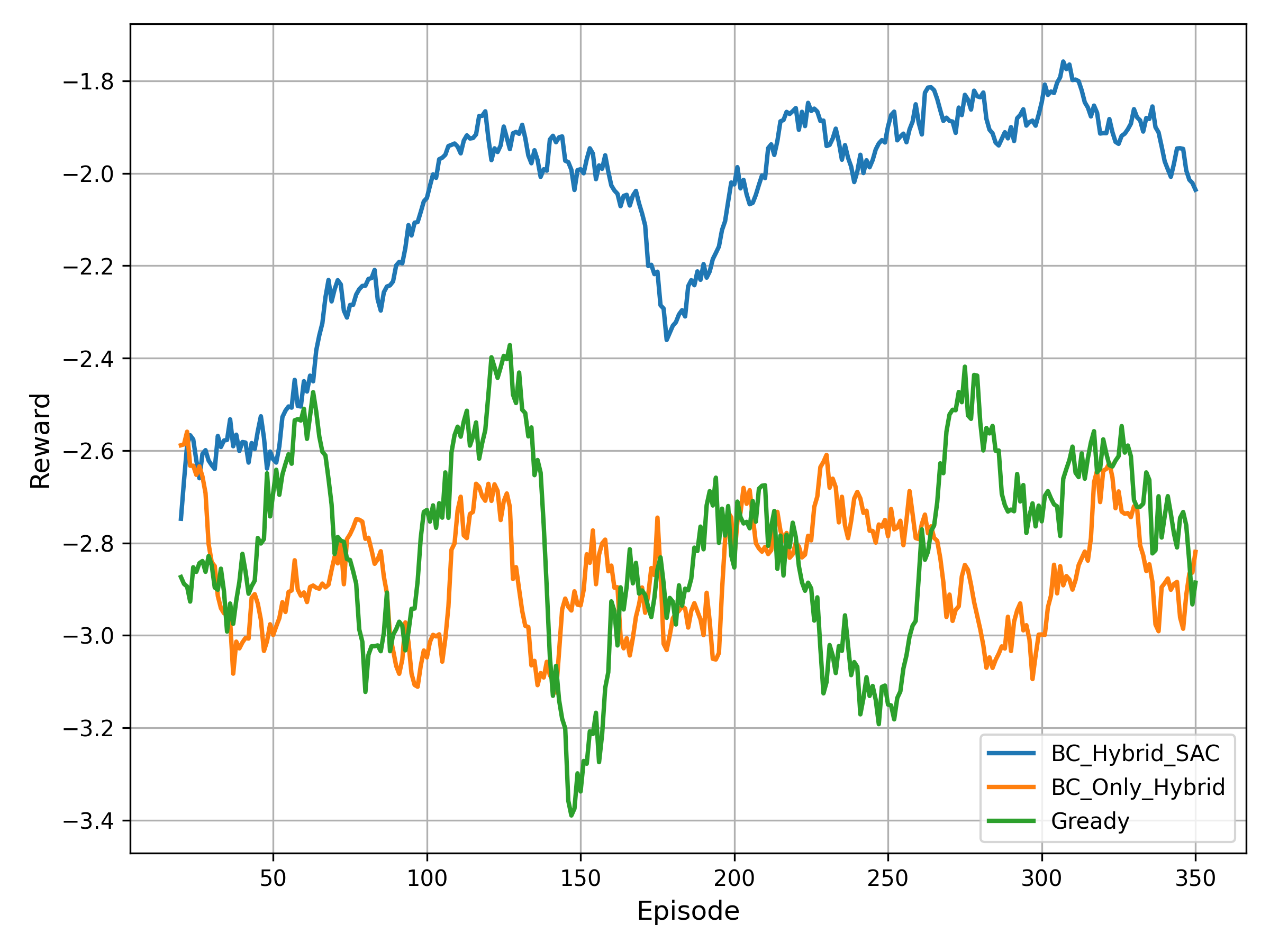}
        \label{Compare_RL_node18}
    }
    \subfigure[$\#$Nodes = 20]{
        \includegraphics[width=0.46\linewidth]{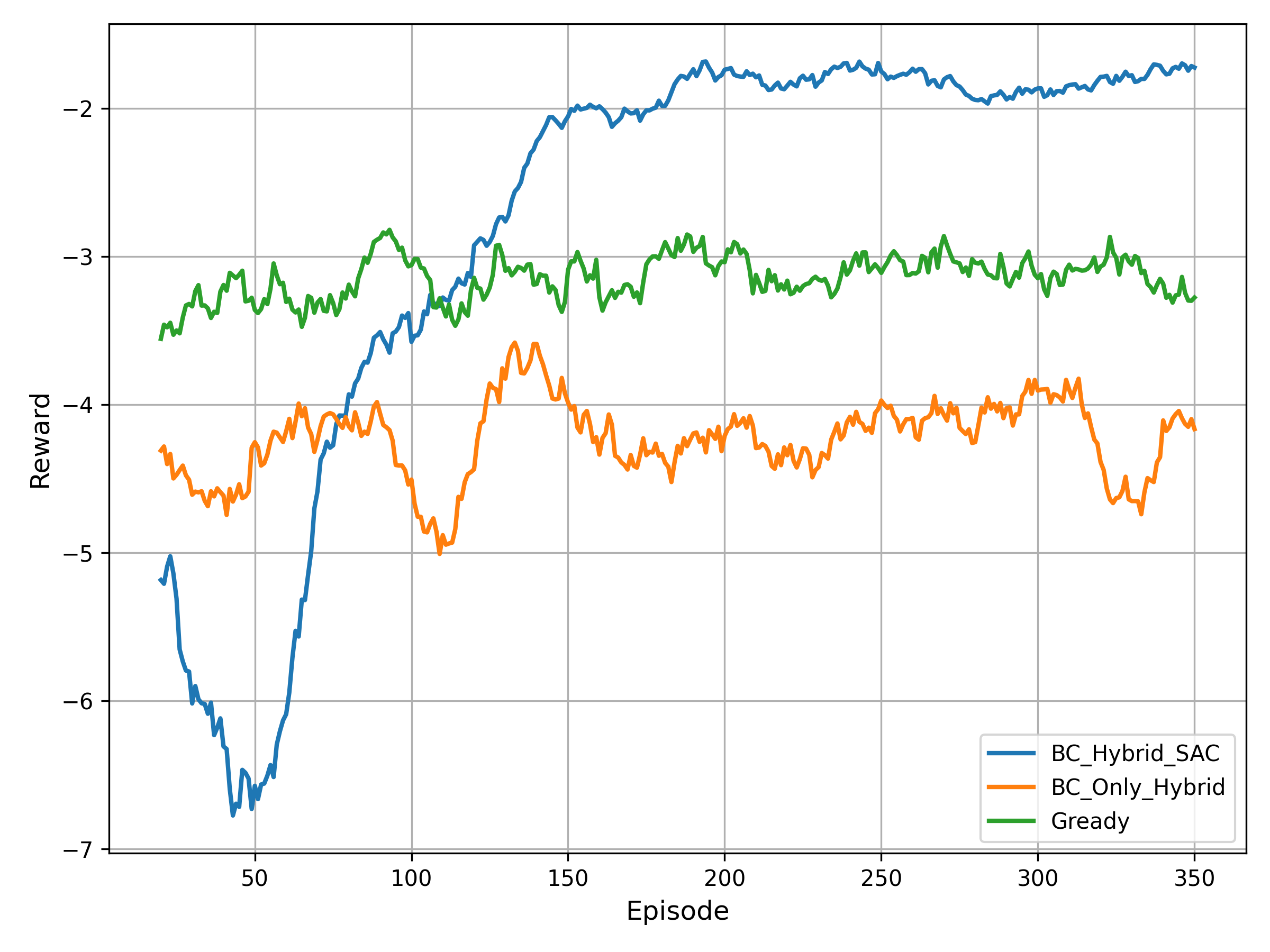}
        \label{Compare_RL_node20}
    }
    \caption{Comparison of convergence performance between BC-only and BC+SAC training across different node settings. Each subfigure shows the reward trajectories of BC\_Only\_Hybrid and BC\_Hybrid\_SAC under various node counts.}
    \label{RLrewardcomp}
\end{figure}

\subsubsection{Impact of RL after BC}

To assess the contribution of reinforcement learning after imitation pretraining, we compare the performance of BC\_Only\_Hybrid and BC\_Hybrid\_SAC across different node configurations. Both models are initialized using the same offline expert dataset, which is generated by the rule-based Greedy algorithm. The only difference lies in whether online SAC training is applied after behavior cloning. The comparison results are shown in Fig.~\ref{RLrewardcomp}.

Across all settings, BC\_Hybrid\_SAC consistently achieves higher rewards and faster convergence. In contrast, BC\_Only\_Hybrid tends to plateau near the reward level of the expert policy, indicating that behavior cloning alone reproduces expert behavior but fails to improve upon it. For example, under the 15-node and 18-node settings, BC\_Hybrid\_SAC reaches rewards of approximately $-2.3$ and $-1.8$, while both the BC\_Only\_Hybrid and the expert baseline remain around $-2.9$ and $-3.0$. In the 20-node setting, the gap becomes even more significant, with BC\_Hybrid\_SAC stabilizing near $-2.0$ and the expert remaining below $-4.0$.

These results confirm that while behavior cloning provides a useful warm start, it lacks adaptability in dynamic, high-dimensional environments. Reinforcement learning allows the policy to continuously refine and improve beyond the expert, leading to better final performance and enhanced robustness.

\subsection{Experimental Results: Detailed Metric Comparison}

\begin{figure}[!t]
    \centering
    \subfigure[Total Complete Time]{
        \includegraphics[width=0.46\linewidth]{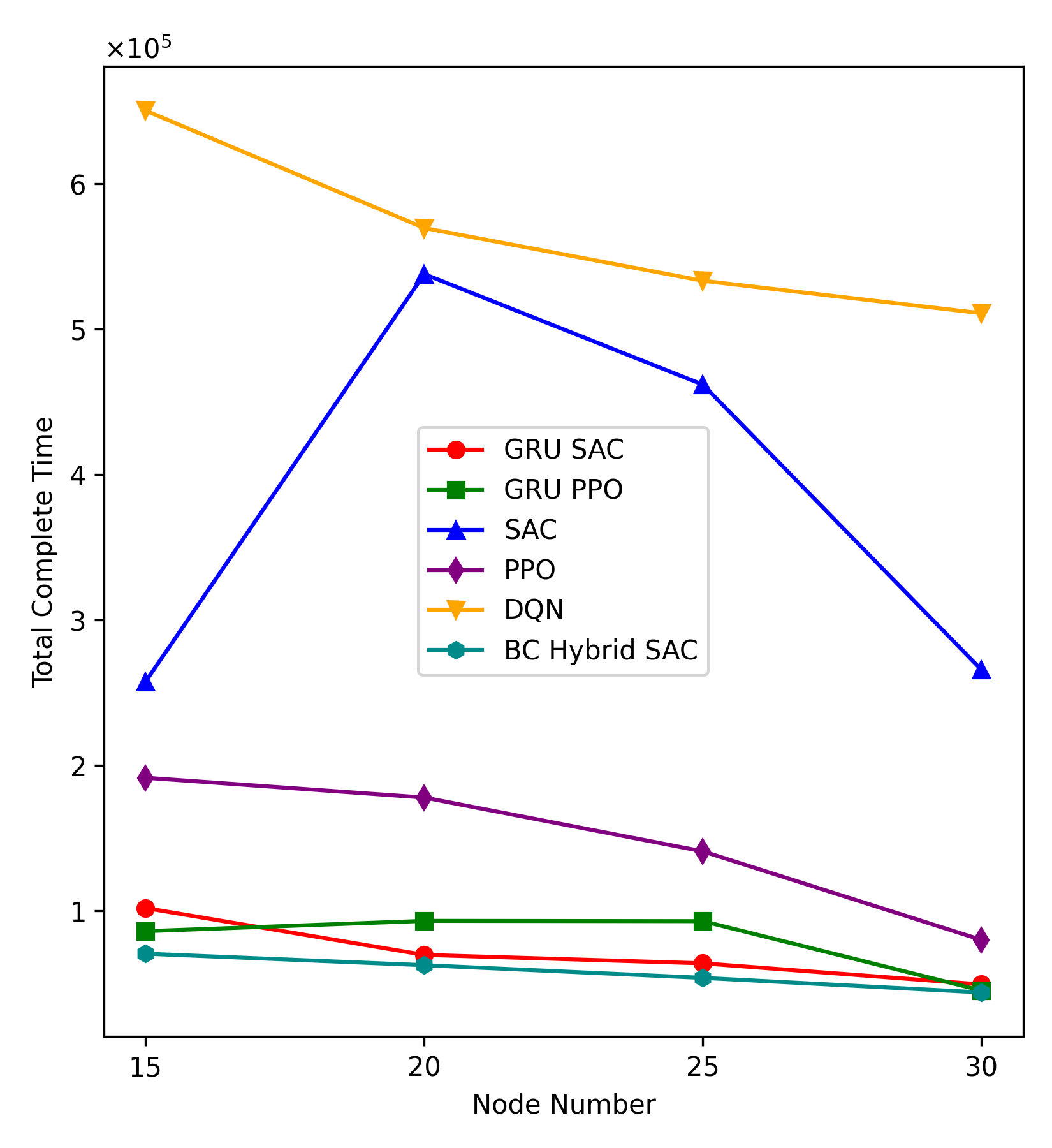}
        \label{Time_node}
    }
    \subfigure[Total Energy Consumption]{
        \includegraphics[width=0.46\linewidth]{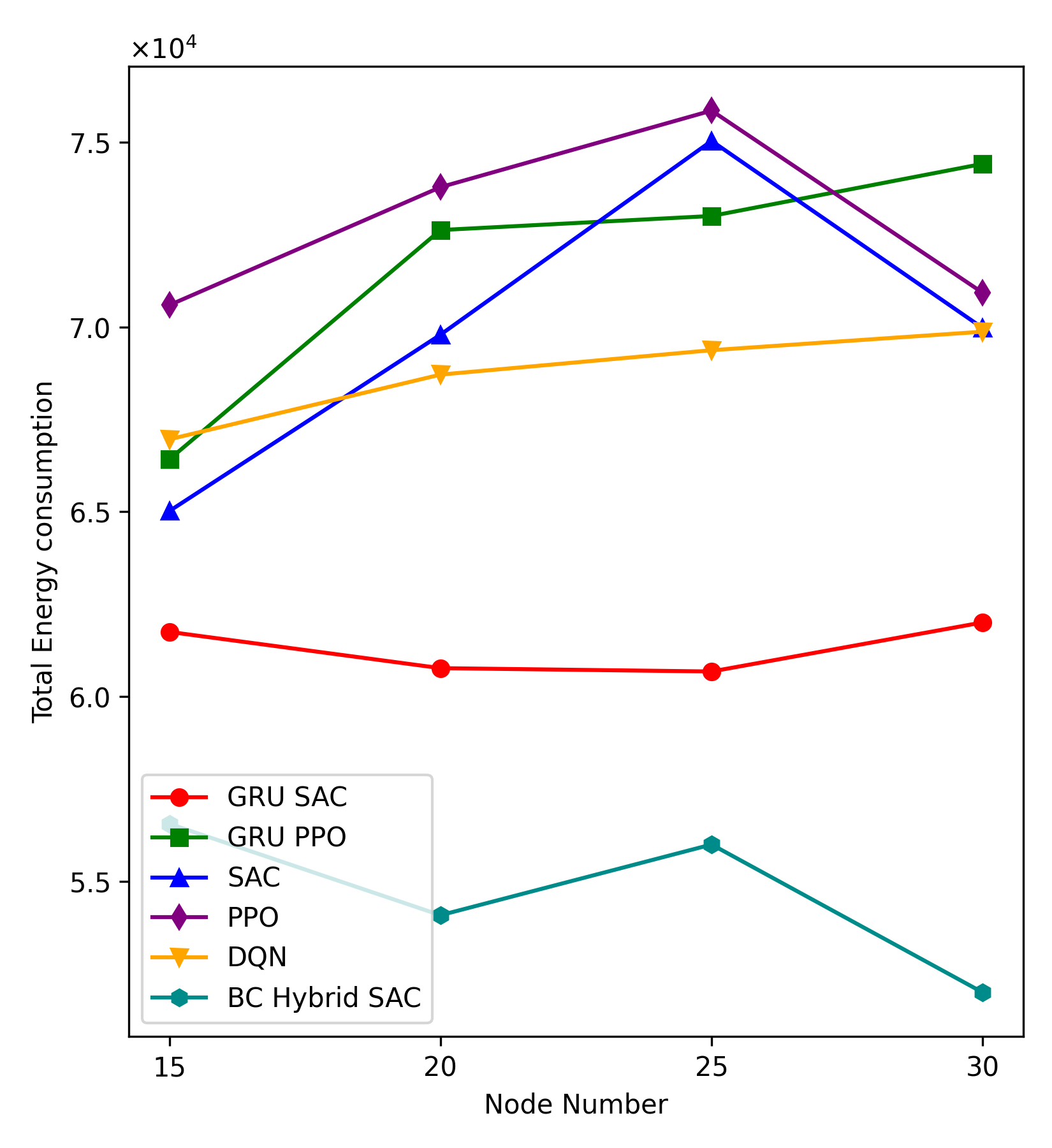}
        \label{Energy_node}
    }

    \subfigure[Image Download Time]{
        \includegraphics[width=0.46\linewidth]{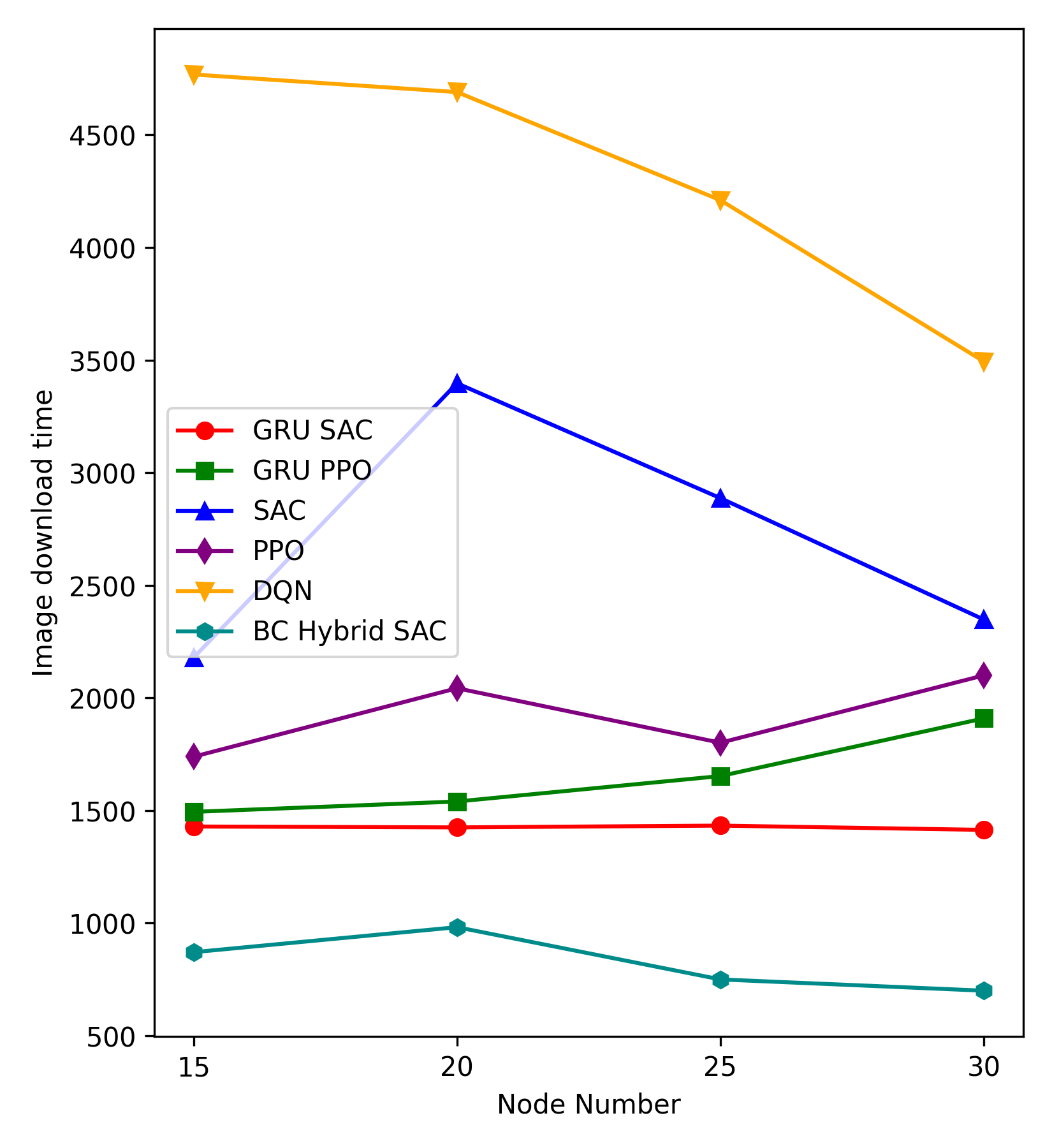}
        \label{Image_node}
    }
    \subfigure[On-time Completion Rate]{
        \includegraphics[width=0.46\linewidth]{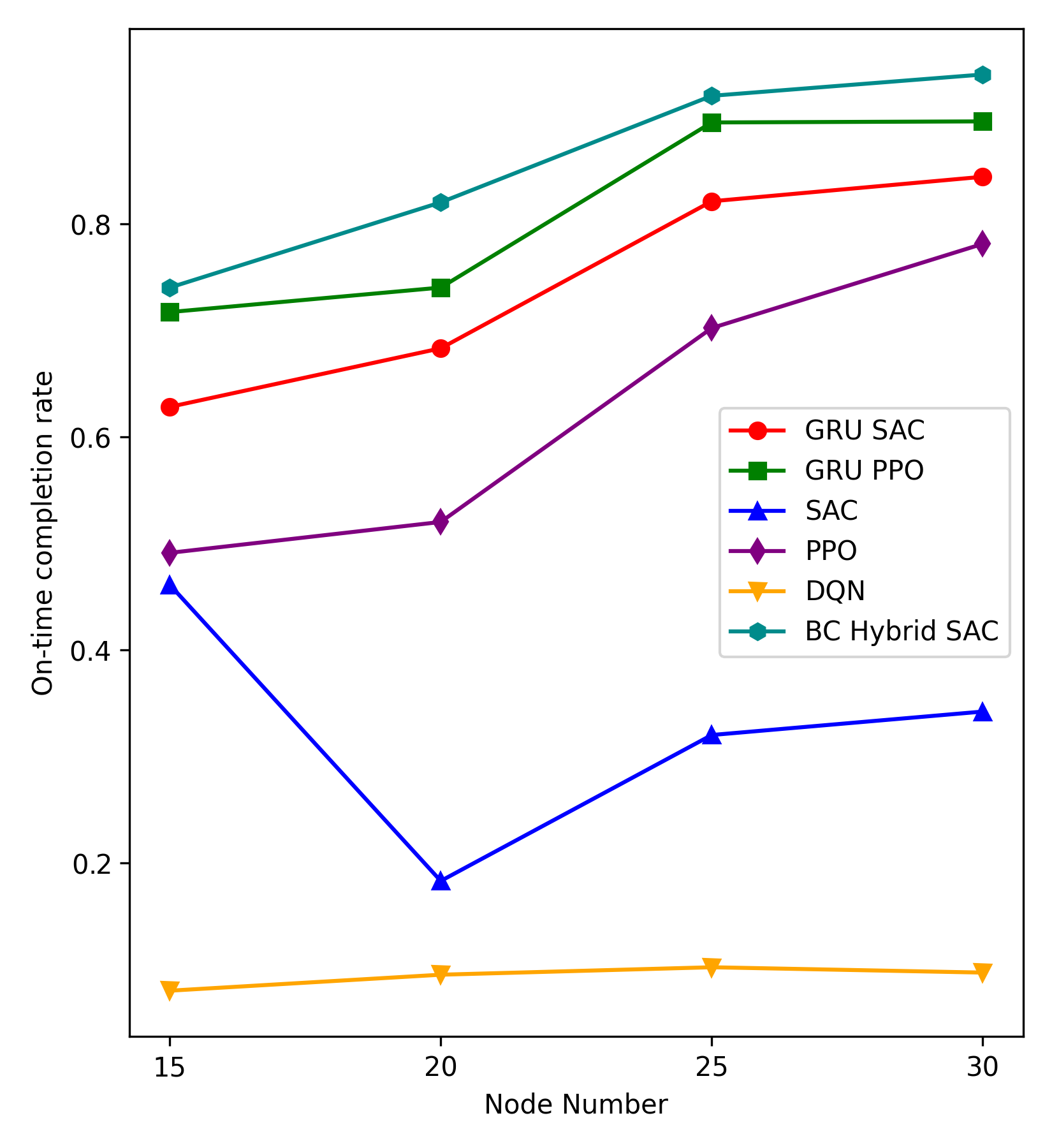}
        \label{Complete_node}
    }
    \caption{Results with the varying number of nodes in different algorithms.}
    \label{node}
\end{figure}

\subsubsection{Performance with the different number of nodes} 
We show the impact of the number of nodes on total completion time, total energy consumption, image download time, and on-time completion ratio in Fig.~\ref{node}. As shown in Fig.~\ref{Time_node}, BC\_Hybrid\_SAC achieves the lowest total completion time across all settings, and its advantage becomes more evident as the number of nodes increases. Although the delay performance of GRU\_SAC and GRU\_PPO is comparable, BC\_Hybrid\_SAC significantly reduces energy consumption, as shown in Fig.~\ref{Energy_node}, and consistently outperforms all other methods. In terms of image download latency, BC\_Hybrid\_SAC also achieves the best performance by minimizing redundant transfers, especially in small-node scenarios where resource contention is more likely. As shown in Fig.~\ref{Complete_node}, the on-time completion ratio improves with more nodes, and BC\_Hybrid\_SAC maintains the highest rate across all configurations, offering a better trade-off between timeliness and energy efficiency.

\begin{figure}[!t]
    \subfigure[Total Complete Time]{
        \includegraphics[width=0.46\linewidth]{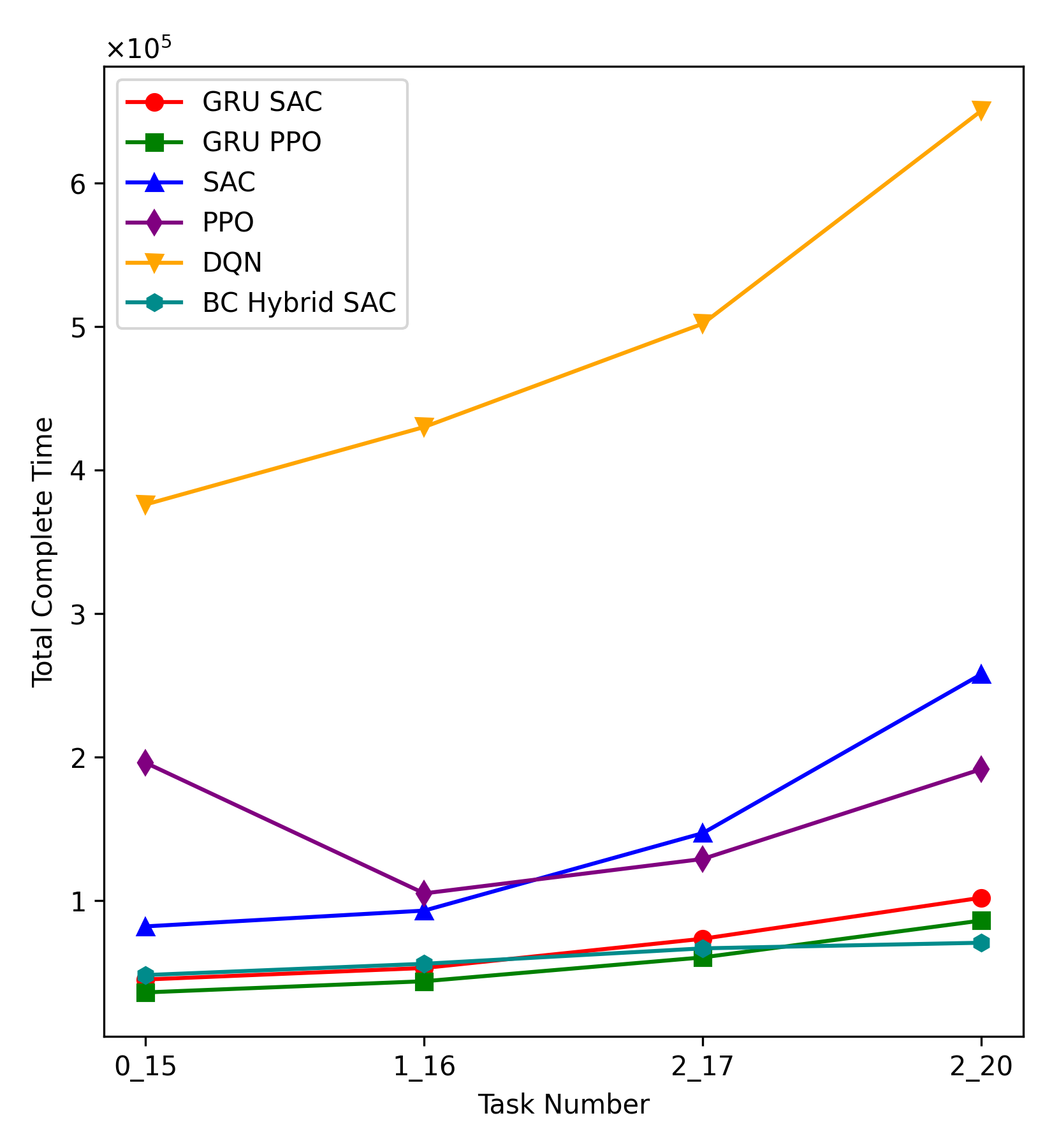}
        \label{Time_task}
    }
    \subfigure[Total Energy Consumption]{
        \includegraphics[width=0.46\linewidth]{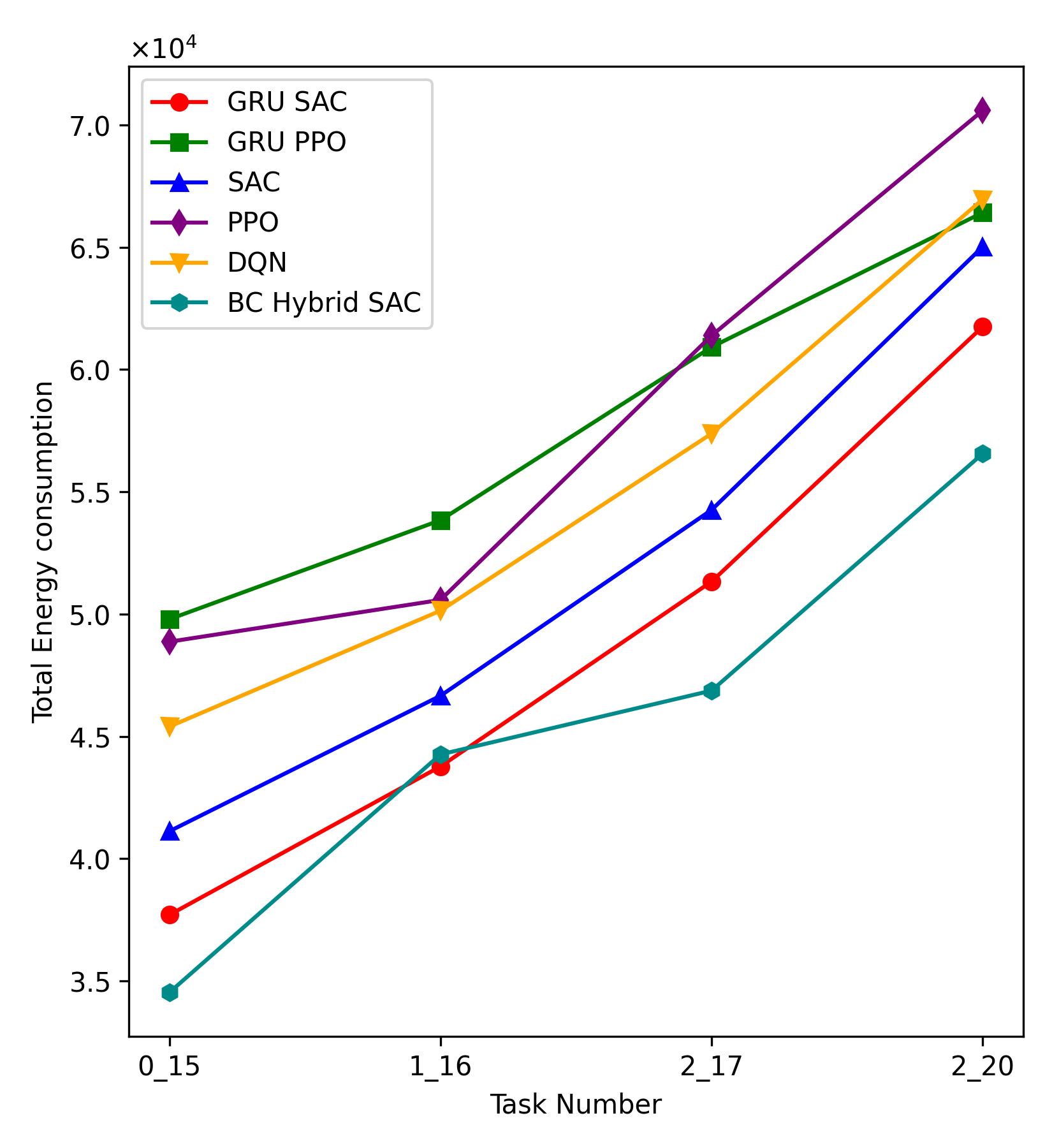}
        \label{Energy_task}
    }

    \subfigure[Image Download Time]{
        \includegraphics[width=0.46\linewidth]{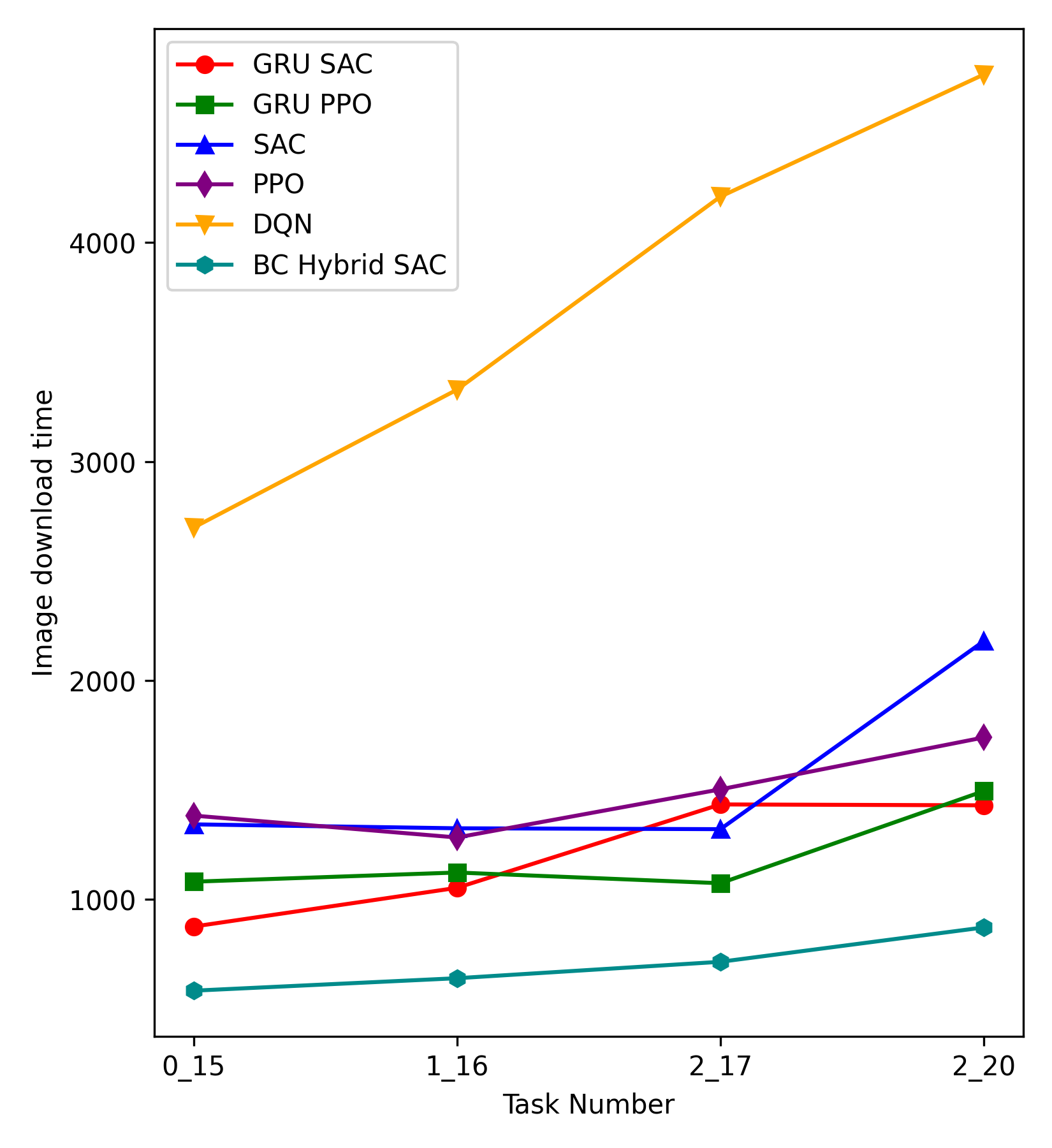}
        \label{Image_task}
    }
    \subfigure[On-time Completion Rate]{
        \includegraphics[width=0.46\linewidth]{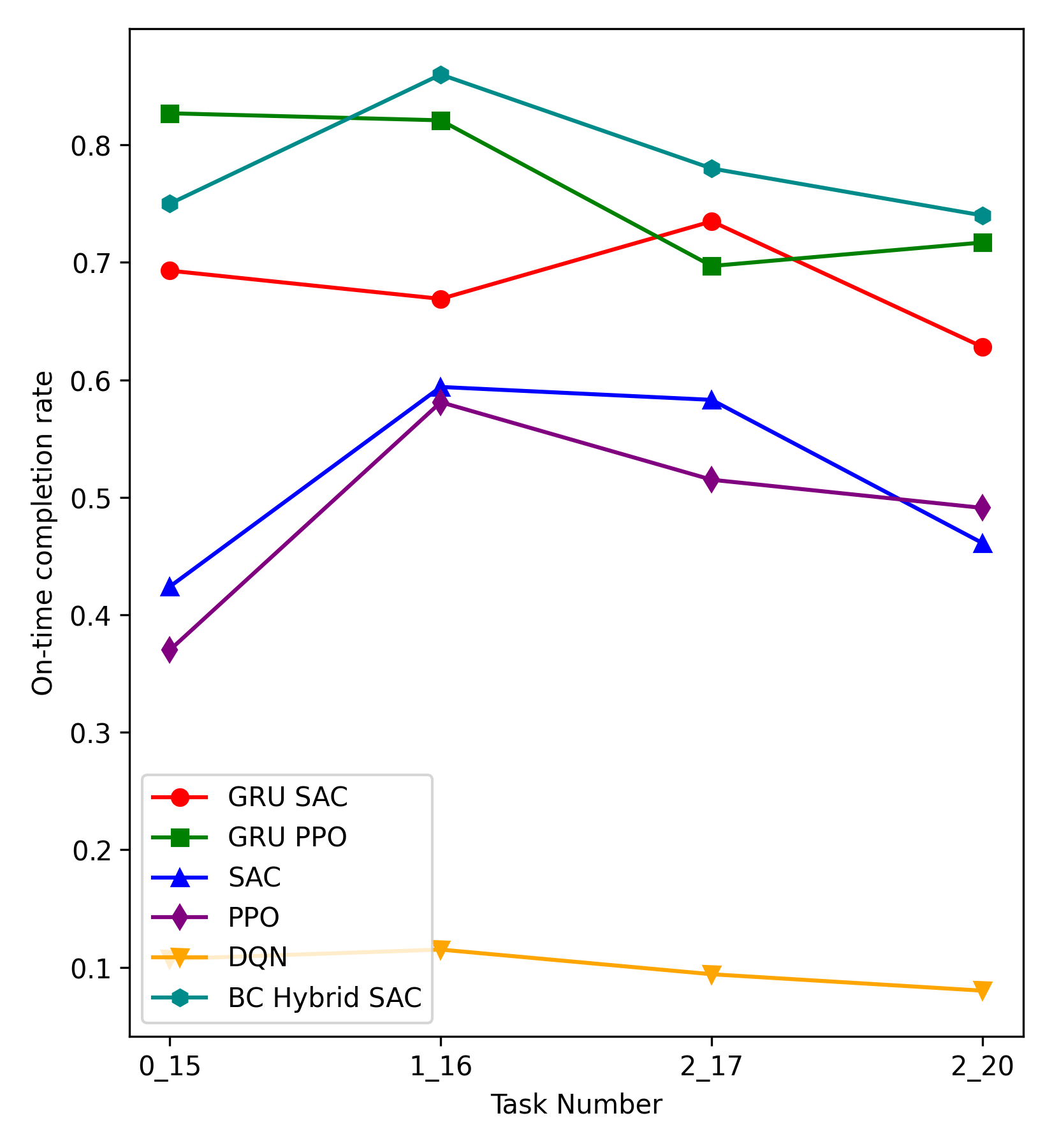}
        \label{Complete_task}
    }
    \caption{Results with the varying number of tasks in different algorithms.}
    \label{task}
\end{figure}

\subsubsection{Performance with Different Numbers of Services} 
Fig.~\ref{task} depicts the impact of varying the number of service requests per time slot, with microservice size fixed at 100 MB. As shown in Fig.~\ref{Time_task}, the total completion time increases as the task number grows, but BC\_Hybrid\_SAC maintains the lowest delay across all settings, significantly outperforming GRU\_SAC and other baselines. In Fig.~\ref{Energy_task}, BC\_Hybrid\_SAC also demonstrates the best energy efficiency, especially under high request volumes. For image download latency (Fig.~\ref{Image_task}), BC\_Hybrid\_SAC again achieves the lowest delay by learning to avoid redundant image transfers. As the number of services increases, the on-time completion ratio decreases, as shown in Fig.~\ref{Complete_task}. While GRU\_PPO slightly outperforms GRU\_SAC in low-load settings, BC\_Hybrid\_SAC consistently achieves the highest completion ratio overall, indicating better adaptability and robustness under dynamic workloads.

\begin{figure}[!t]
    \centering
    \subfigure[Total Complete Time]{
        \includegraphics[width=0.46\linewidth]{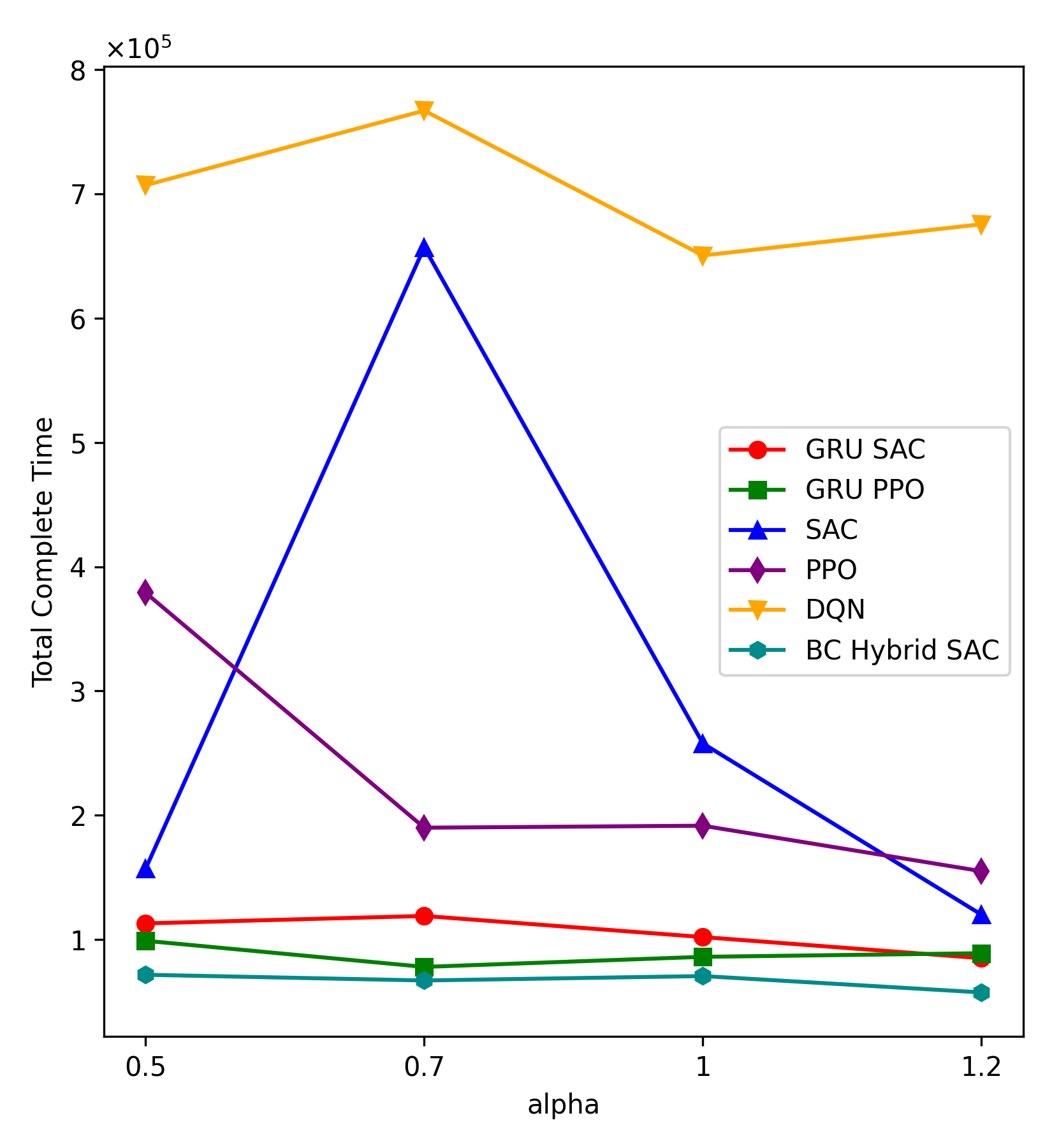}
        \label{Time_alpha}
    }
    \subfigure[Total Energy Consumption]{
        \includegraphics[width=0.46\linewidth]{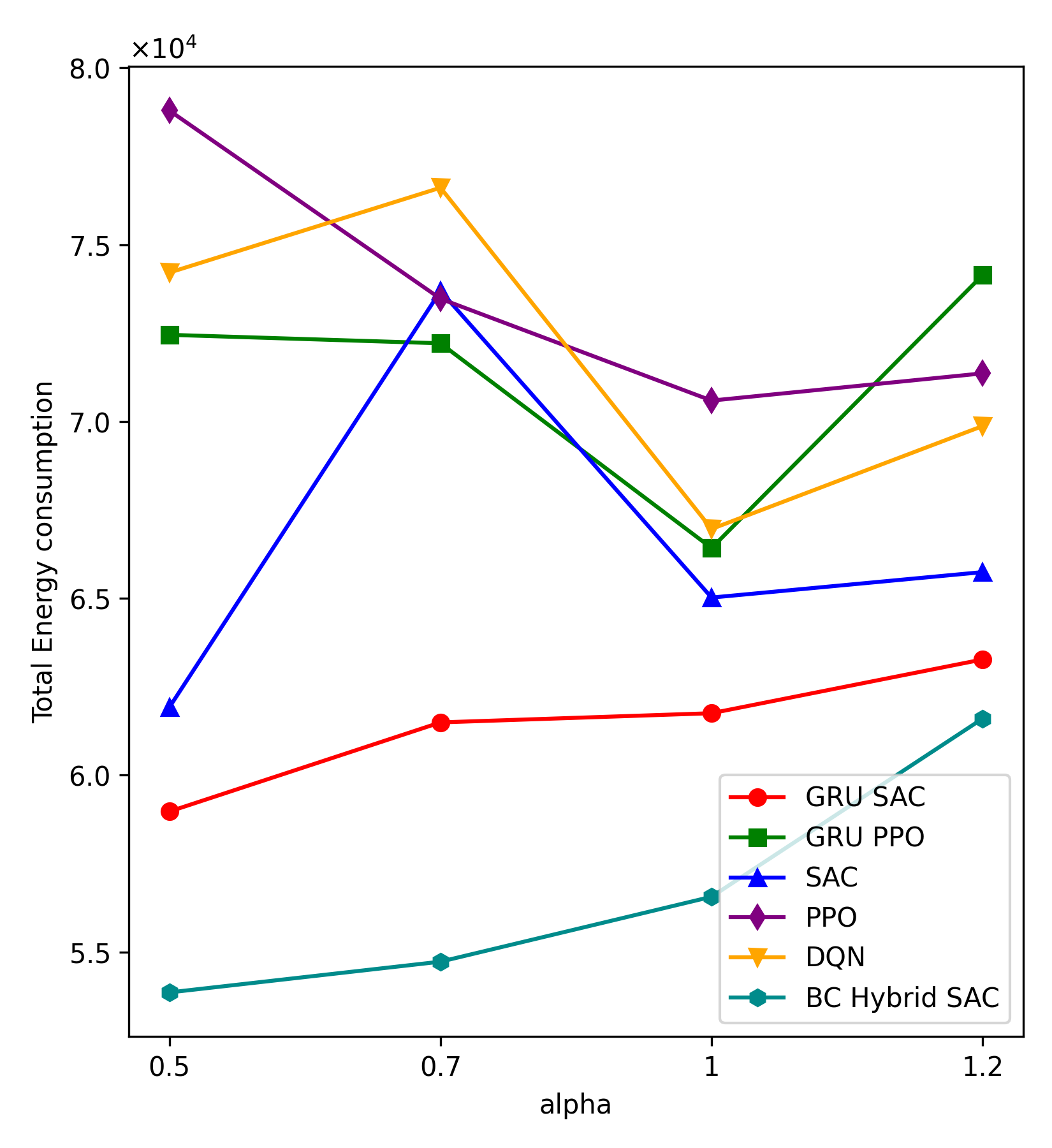}
        \label{Energy_alpha}
    }

    \subfigure[Image Download Time]{
        \includegraphics[width=0.46\linewidth]{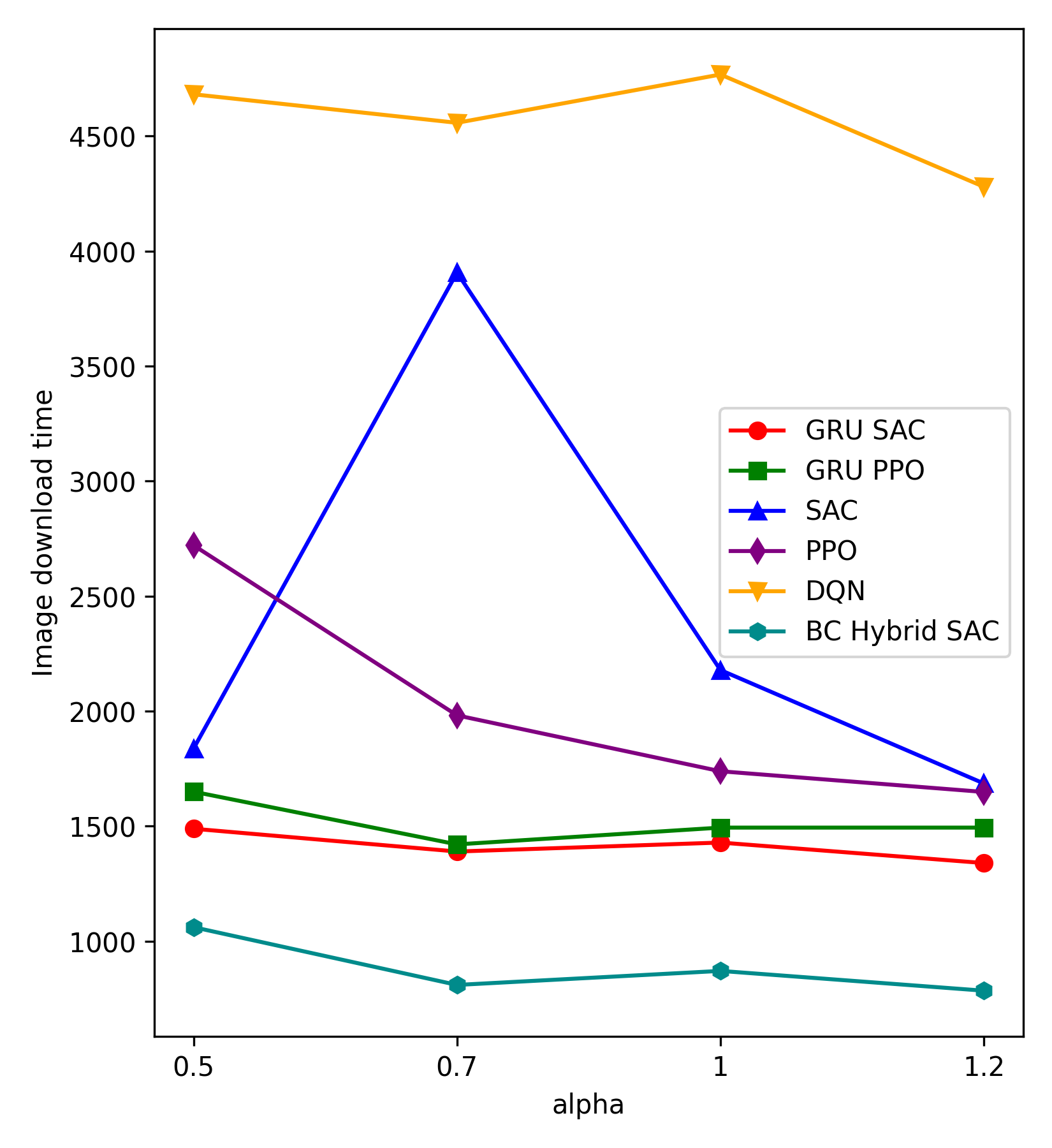}
        \label{Image_alpha}
    }
    \subfigure[On-time Completion Rate]{
        \includegraphics[width=0.46\linewidth]{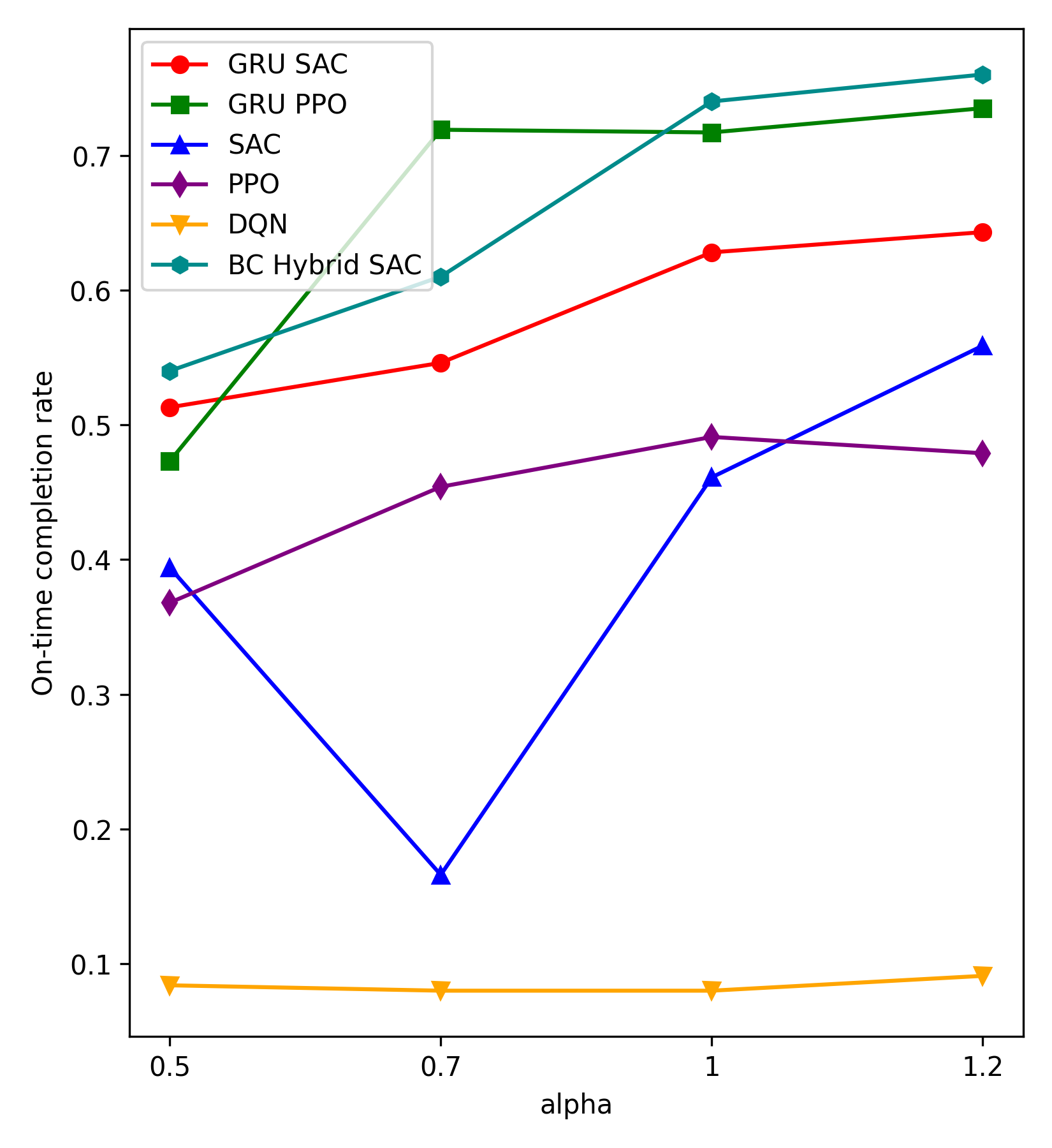}
        \label{Complete_alpha}
    }
    \caption{Results with varying $\alpha$ in the reward in different algorithms.}
    \label{alpha}
\end{figure}

\subsubsection{Performance with Different $\alpha$ in the Reward Function} 
We show the impact of varying $\alpha$ values in Fig.~\ref{alpha}, where $\alpha$ controls the trade-off between latency and energy consumption. As shown in Fig.~\ref{Time_alpha} and Fig.~\ref{Energy_alpha}, increasing $\alpha$ leads to lower total completion time but higher energy usage. GRU\_PPO initially achieves higher on-time completion rates by stabilizing early on delay-focused policies, but it lacks sufficient exploration to optimize energy efficiency. GRU\_SAC performs better by maintaining entropy-driven exploration throughout training. Notably, BC\_Hybrid\_SAC consistently achieves the best balance across all $\alpha$ values, offering lower latency, reduced energy consumption, and higher on-time completion ratios as shown in Fig.~\ref{Complete_alpha}.

\section{Conclusion}\label{sec6}
In this paper, we consider an online container-based microservice scheduling problem with dynamic computing power, aiming to minimize the total service delay and energy consumption. We first model the edge environment with resource constraints and define a scheduling objective that jointly considers latency, energy, and feasibility. To improve the learning stability under cold-start conditions, we introduce a two-phase learning framework that combines behavior cloning with a GRU-enhanced SAC RL framework. An enhanced policy network is designed to separately encode node-level and task-level features, enabling sequential and resource-aware scheduling. Experimental results show that the proposed method achieves faster convergence and better performance compared to existing baselines, 
improving the total objective by $50\%$ and convergence speed by $70\%$ as well as maintaining high stability under different edge configurations.

\section*{Acknowledgment}
This work was supported in part by the National Natural Science Foundation of China (NSFC) under Grant No. 62202055, the Guangdong Basic and Applied Basic Research Foundation under Grant No. 2025A1515012843, the Start-up Fund from Beijing Normal University under Grant No. 312200502510, the Internal Fund from Beijing Normal-Hong Kong Baptist University under Grant No. UICR0400003-24, the Project of Young Innovative Talents of Guangdong Education Department under Grant No. 2022KQNCX102, and the Interdisciplinary Intelligence SuperComputer Center of Beijing Normal University (Zhuhai).

\ifCLASSOPTIONcaptionsoff
  \newpage
\fi

\bibliographystyle{IEEEtran}
\bibliography{references}

\begin{IEEEbiography}[{\includegraphics[width=1in,height=1.25in,clip,keepaspectratio]{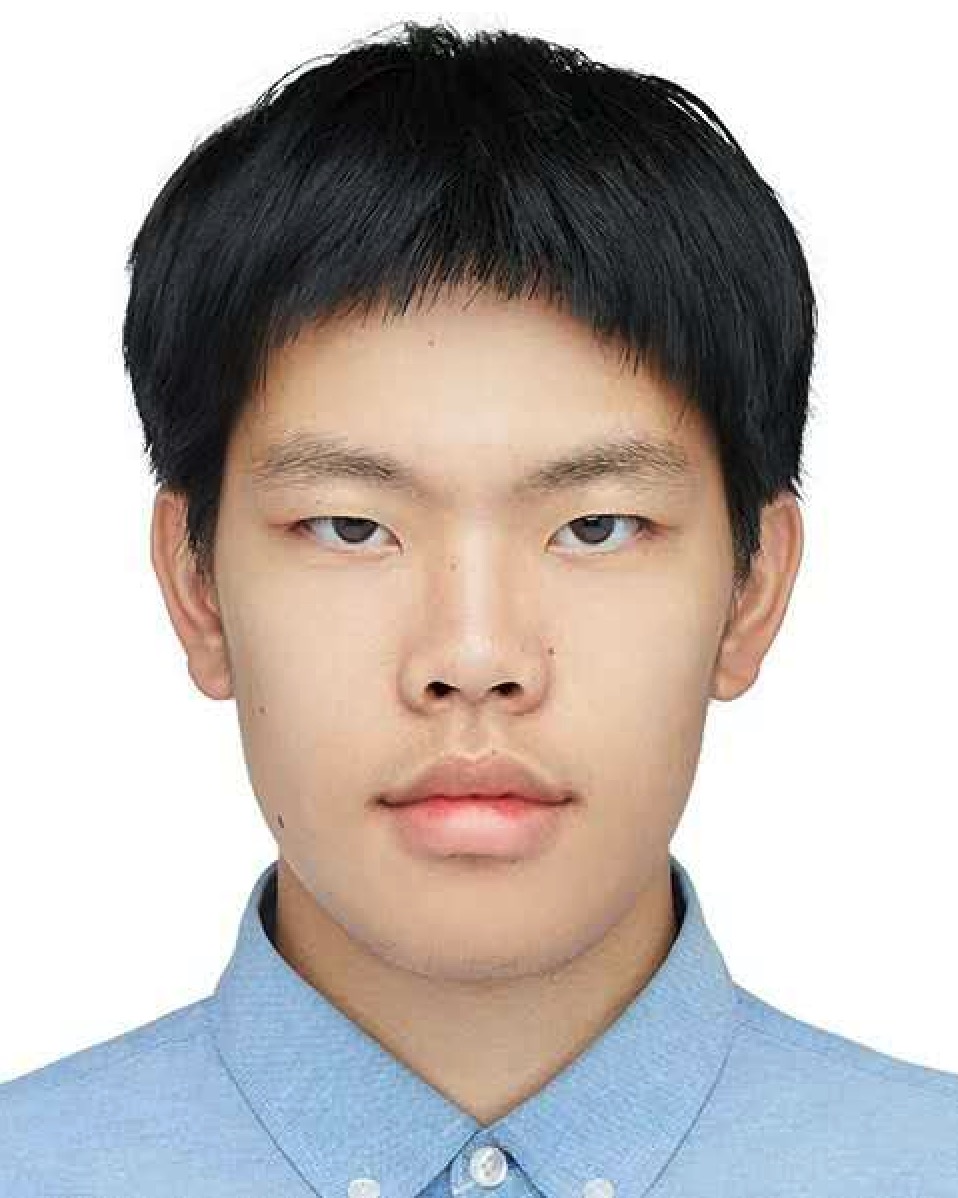}}]{Jingxi Lu} received the B.S. degree in computer science and technology from Beijing Normal-Hong Kong Baptist University, Zhuhai, China. He is currently pursuing the M.S. degree in electrical and computer engineering at the University of Southern California, Los Angeles, CA, USA. His research interests include reinforcement learning, robotics, edge computing, and machine learning.
    
\end{IEEEbiography}

\begin{IEEEbiography}[{\includegraphics[width=1in,height=1.25in,clip,keepaspectratio]{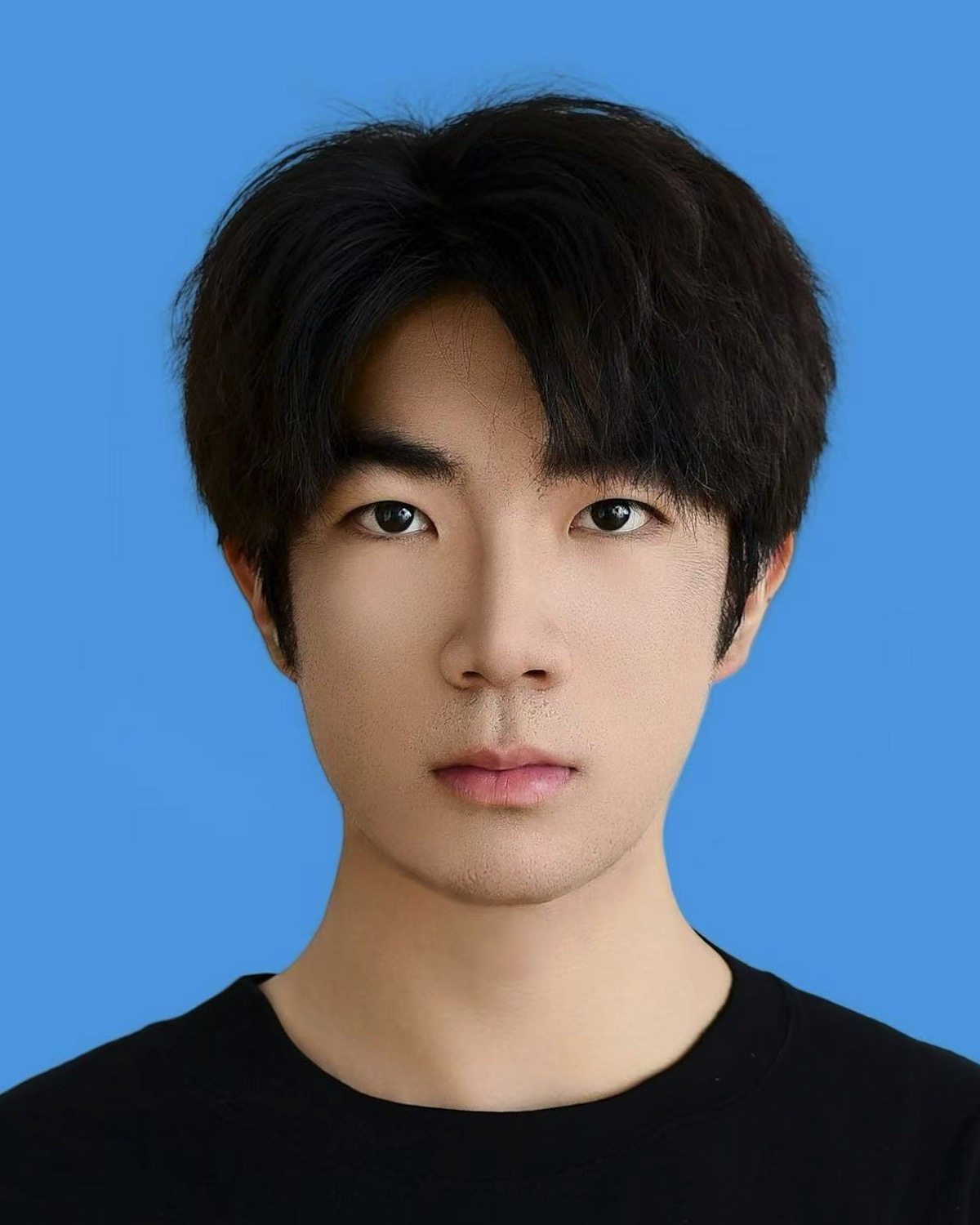}}]{Wenhao Li}received the B.S. degree in computer science and technology from Beijing Normal-Hong Kong Baptist University, Zhuhai, China. He is currently pursuing the M.S. degree in data science at the University of Sydney, Sydney, NSW, Australia. His research interests include machine learning, deep learning, and data analytics.
    
\end{IEEEbiography}

\begin{IEEEbiography}[{\includegraphics[width=1in,height=1.25in,clip,keepaspectratio]{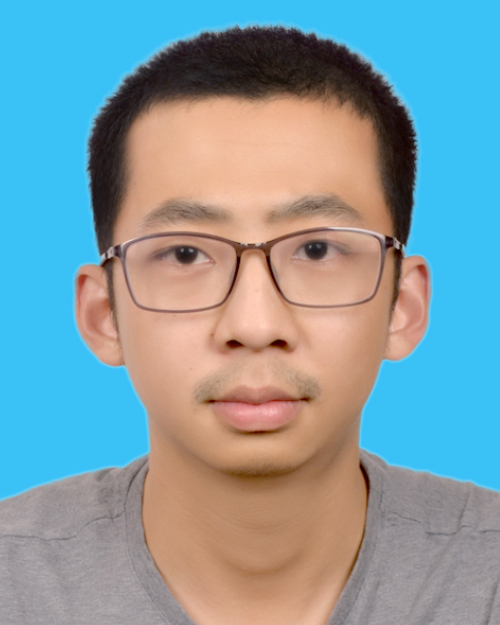}}]{Jianxiong Guo}
    (Member, IEEE) received his Ph.D. degree from the Department of Computer Science, University of Texas at Dallas, USA, in 2021, and his B.E. degree from the School of Chemistry and Chemical Engineering, South China University of Technology, China, in 2015. He is currently an Associate Professor with the Advanced Institute of Natural Sciences, Beijing Normal University, and also with the Guangdong Key Lab of AI and Multi-Modal Data Processing, Beijing Normal-Hong Kong Baptist University, Zhuhai, China. He is a member of IEEE/ACM/CCF. He has published more than 100 peer-reviewed papers and has been the reviewer for many famous international journals/conferences. His research interests include social networks, wireless sensor networks, combinatorial optimization, and machine learning.
\end{IEEEbiography}

\begin{IEEEbiography}[{\includegraphics[width=1in,height=1.25in,clip,keepaspectratio]{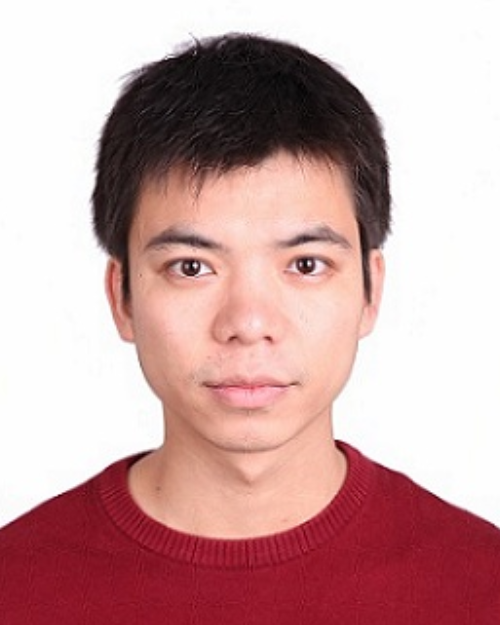}}]{Xingjian Ding}
    received the B.E. degree in electronic information engineering from Sichuan University, Chengdu, China, in 2012, the M.S. degree in software engineering from Beijing Forestry University, Beijing, China, in 2017, and the Ph.D. degree from the School of Information, Renmin University of China, Beijing, in 2021. He is currently an Assistant Professor with the School of Software Engineering, Beijing University of Technology, Beijing. His research interests include wireless rechargeable sensor networks, approximation algorithms design and analysis, and blockchain.
\end{IEEEbiography}

\begin{IEEEbiography}[{\includegraphics[width=1in,height=1.25in,clip,keepaspectratio]{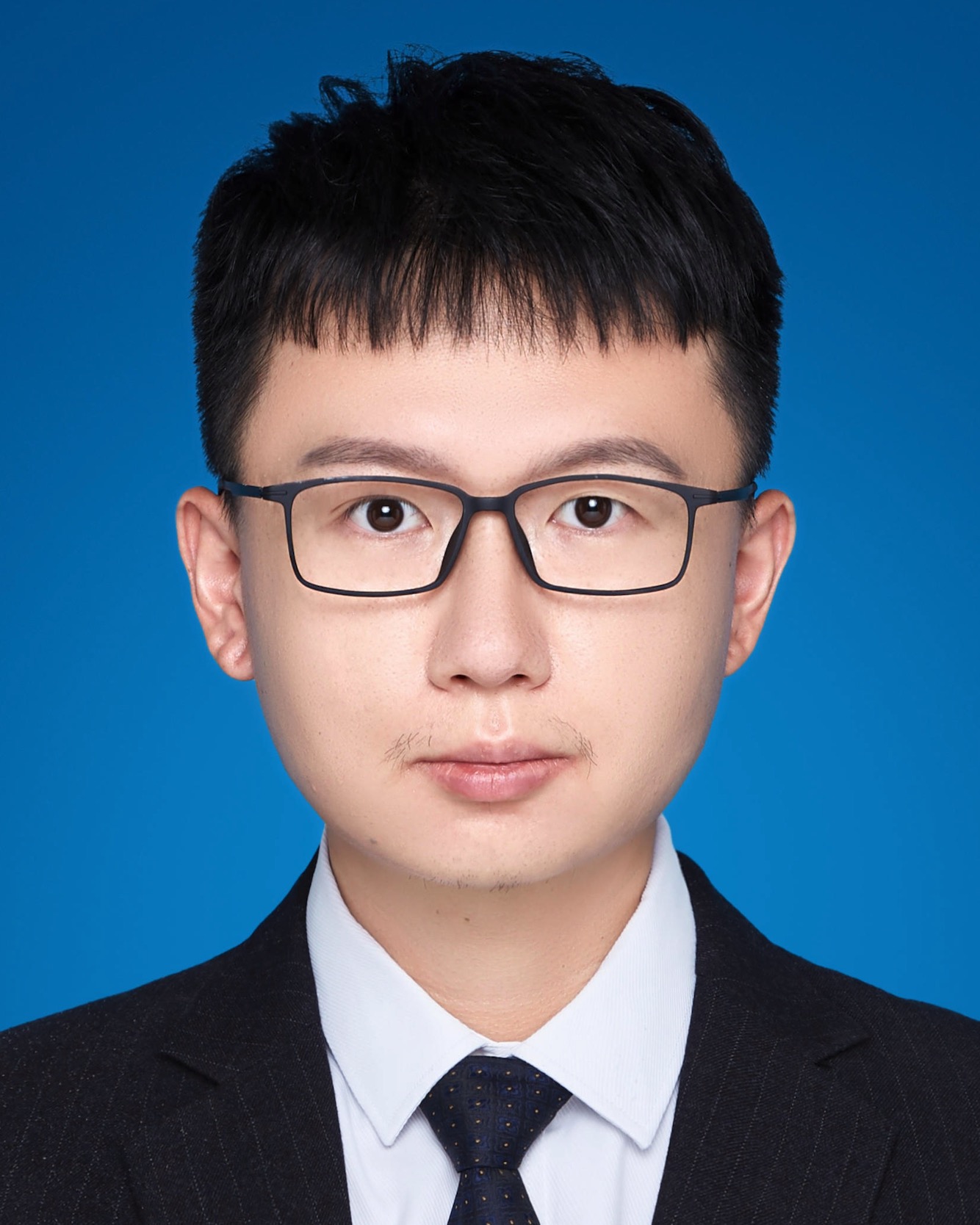}}]{Zhiqing Tang}
    (Member, IEEE) received the B.S. degree from the School of Communication and Information Engineering, University of Electronic Science and Technology of China, China, in 2015 and the Ph.D. degree from the Department of Computer Science and Engineering, Shanghai Jiao Tong University, China, in 2022. He is currently an assistant professor with the Institute of Artificial Intelligence and Future Networks, Beijing Normal University, China. His current research interests include edge computing, resource scheduling, and reinforcement learning.
\end{IEEEbiography}

\begin{IEEEbiography}[{\includegraphics[width=1in,height=1.25in,clip,keepaspectratio]{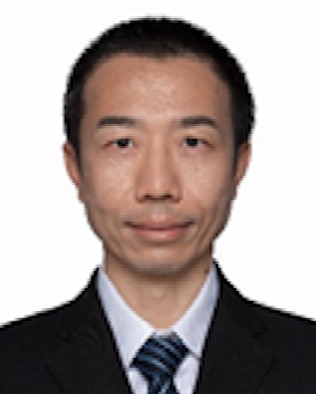}}]{Tian Wang}
    (Senior Member, IEEE) received his BSc and MSc degrees in Computer Science from Central South University in 2004 and 2007, respectively. He received his Ph.D. degree in Computer Science from the City University of Hong Kong in 2011. Currently, he is a professor at the Institute of Artificial Intelligence and Future Networks, Beijing Normal University. His research interests include Internet of Things, edge computing, and mobile computing. He has 30 patents and has published more than 200 papers in high-level journals and conferences. He was a co-recipient of the Best Paper Runner-up Award of IEEE/ACM IWQoS 2024. He has more than 16000 citations, according to Google Scholar. His H-index is 74.
\end{IEEEbiography}

\begin{IEEEbiography}[{\includegraphics[width=1in,height=1.25in,clip,keepaspectratio]{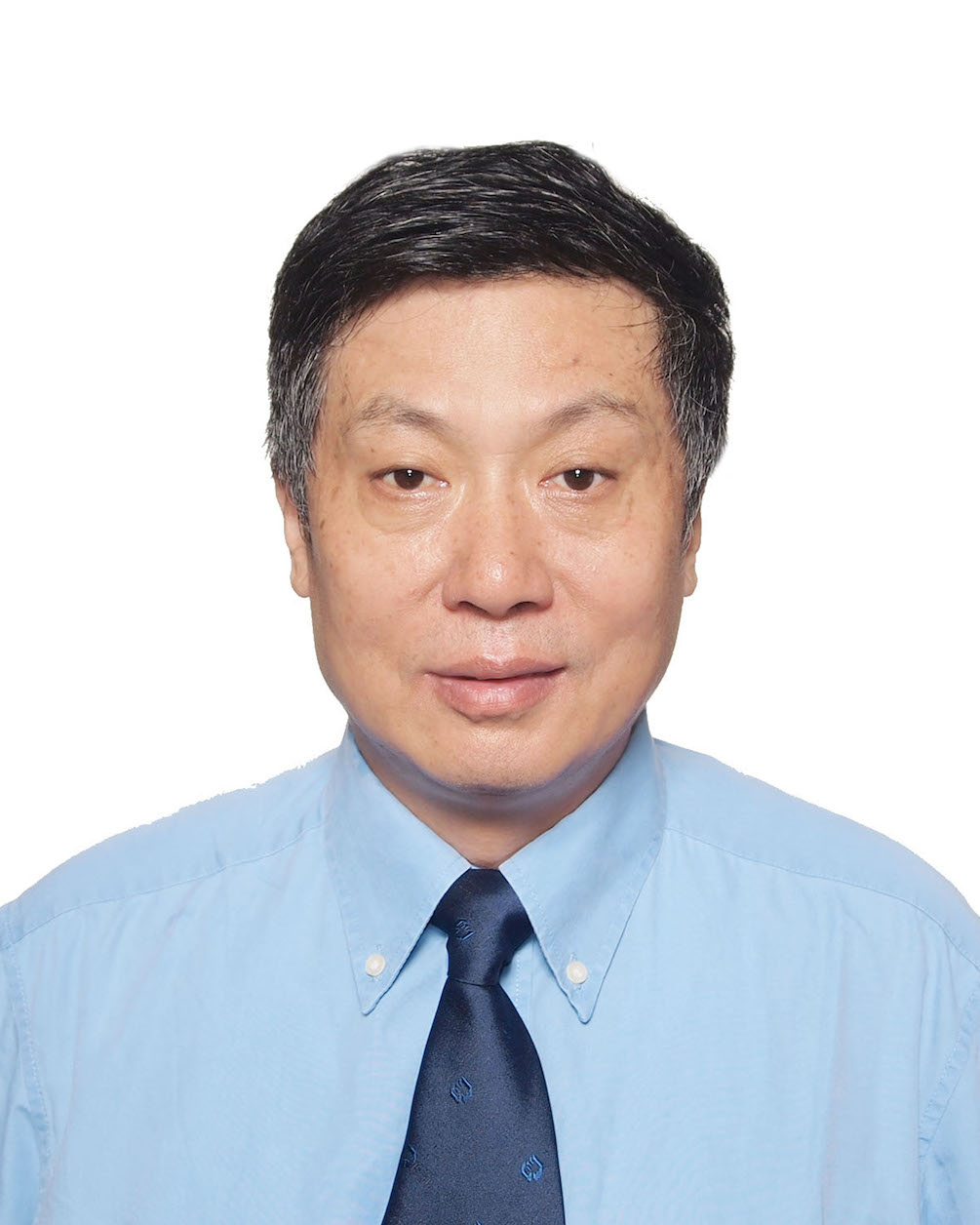}}]{Weijia Jia} (Fellow, IEEE)
    is currently a Chair Professor, Director of the Institute of AI and Future Networks, Beijing Normal University and VP for Research at BNU-HKBU United International College and has been the Chair Professor of Shanghai Jiao Tong University, China. He was the Chair Professor and the Deputy Director of State Kay Laboratory of IoT for Smart City at the University of Macau. He received BSc/MSc from Center South University, China in 82/84 and Master of Applied Sci./PhD from Polytechnic Faculty of Mons, Belgium in 92/93, respectively, all in computer science. From 93-95, he joined German National Research Center for Information Science (GMD) in Bonn (St. Augustine) as a research fellow. From 95-13, he worked in City University of Hong Kong as a professor. His contributions have been recognized as optimal network routing and deployment; anycast and QoS routing, sensors networking, AI (knowledge relation extractions; NLP, etc.), and edge computing. He has over 600 publications in the prestige international journals/conferences and research books and book chapters. He has received the best product awards from International Science \& Tech. Expo (Shenzhen) in 20112012 and the 1st Prize of Scientific Research Awards from the Ministry of Education of China in 2017 (list 2). He has served as area editor for various prestige international journals, chair, and PC member/keynote speaker for many top international conferences. He is the Fellow of IEEE and the Distinguished Member of CCF.
\end{IEEEbiography}

\end{document}